\documentclass[twocolumn,aps,pra,preprintnumbers,bibnotes,amsmath,amssymb]{revtex4-1}
\usepackage{graphicx}
\usepackage{bm}
\usepackage{color}
\usepackage{multirow}

\usepackage[colorlinks,urlcolor=blue,citecolor=blue,linkcolor=blue]{hyperref}

\begin{document}

\title{Feshbach resonances, weakly bound molecular states and coupled-channel potentials\\
for cesium at high magnetic fields}

\author{Martin Berninger}
\author{Alessandro Zenesini}
\author{Bo Huang}
\author{Walter Harm}
\author{Hanns-Christoph N\"{a}gerl}
\author{Francesca Ferlaino}
\author{Rudolf Grimm}
\affiliation{Institut f\"ur Experimentalphysik and Zentrum f\"ur Quantenphysik,
Universit\"at Innsbruck, 
6020 Innsbruck, Austria} \affiliation{Institut f\"ur Quantenoptik und
Quanteninformation,
 \"Osterreichische Akademie der Wissenschaften, 6020 Innsbruck,
 Austria}
\author{Paul S. Julienne}
\affiliation{Joint Quantum Institute, NIST and the University of Maryland, 100
Bureau Drive Stop 8423, Gaithersburg, Maryland 20899-8423, USA}
\author{Jeremy M. Hutson}
\affiliation{Joint Quantum Centre (JQC) Durham/Newcastle, Department of
Chemistry, Durham University, South Road, Durham, DH1~3LE, United
Kingdom}

\date{\today}


\begin{abstract}
We explore the scattering properties of ultracold ground-state Cs atoms at
magnetic fields between 450~G (45~mT) and 1000~G. We identify 17 new Feshbach resonances,
including two very broad ones near 549~G and 787~G. We measure the binding
energies of several different dimer states by magnetic field modulation
spectroscopy. We use least-squares fitting to these experimental results,
together with previous measurements at lower field, to determine a new
6-parameter model of the long-range interaction potential, designated M2012.
Coupled-channels calculations using M2012 provide an accurate mapping between
the s-wave scattering length and the magnetic field over the entire range of
fields considered. This mapping is crucial for experiments that rely on precise
tuning of the scattering length, such as those on Efimov physics.
\end{abstract}

\maketitle

\section{Introduction}
Cold cesium atoms have provided the foundation for many important experiments
in basic science  and also find application in precise atomic clocks. A thorough understanding of the collisions and interactions of two Cs
atoms is crucial to interpret such experiments and optimize the applications.
In particular, Cs has a complicated spectrum of magnetically tunable Feshbach
resonances \cite{Chin2010fri}, which allow precise control of the two-body
interactions. These resonances are due to near-threshold bound states of the
diatomic molecule Cs$_2$ that can be tuned to match the near-zero energy of the
colliding atoms. The resonances also allow an atomic sample to be converted
with high efficiency into diatomic molecules by tuning an applied magnetic
field across a resonance \cite{Herbig2003poa}. We have previously shown that the weakly bound
molecules formed in this way can be converted into deeply bound
\cite{Danzl2008qgo} or even ground-state \cite{Danzl2010auh} molecules by
stimulated Raman adiabatic passage.

Early work with Cs atoms and its interactions was at low magnetic
field, $B \lesssim 250$~G \cite{Note1}. Results on low-field Feshbach resonances \cite{Vuletic1999ool,
Chin2000hrf, Chin2003sdo, Chin2004pfs} made it possible to construct
theoretical models of the near-threshold bound and scattering states of
two cold Cs atoms \cite{Leo2000cpo, Chin2004pfs}. These models used the
full Hamiltonian of Cs$_2$, including the potential energy curves of
the $^1\Sigma_g^+$ singlet and $^3\Sigma_u^+$ triplet states, the
molecular spin-spin interaction, and the atomic hyperfine interactions.
Fitting the data allowed four key parameters of the model to be
adjusted so that the resonance structure could be reproduced accurately
in the low-field region. These early models yielded an understanding of
the large clock shifts in Cs atomic fountain clocks \cite{Leo2001cfs}, the
anomalously large loss rates for collisions of doubly spin-polarized Cs
atoms \cite{Soding1998gsr}, and the magnetic field regions of moderate positive
scattering length where Bose-Einstein condensation (BEC) was
possible~\cite{Weber2003bec}. Subsequent measurements of the binding energies of
weakly bound dimer states at fields up to 60~G \cite{Mark2007sou,
Knoop2008mfm} were mostly in good agreement with calculations based on
the model of Ref.\,\cite{Chin2004pfs}, which we designate M2004.

Ultracold Cs is particularly important for the study of Efimov states
\cite{Efimov1970ela}, which are high-lying bound states of triatomic molecules
that appear when the two-body interaction has a bound state very close to
threshold. Efimov states cause additional loss features close to 2-body
Feshbach resonances. The first observation of an Efimov resonance in Cs
\cite{Kraemer2006efe} was at a field near 8~G. We have recently extended this
to observe Efimov features associated with three additional 2-body Feshbach
resonances, at fields up to 900~G, and reached the remarkable conclusion that
the Efimov features all occur at almost the same 2-body scattering length $a$,
and thus all have almost the same binding energy at unitarity ($a=\infty$)
\cite{Berninger2011uot}. This universality of Efimov states was quite
unexpected, and contradicted previous theoretical indications
\cite{Dincao2009tsr}, although subsequent theoretical work is now starting to
explain its origins \cite{Wang2012oot, Naidon2012poo, Schmidt2012epb,
Chin2011uso}. Efimov states in other ultracold systems are now also being found
to show the same universal behavior \cite{Pollack2009uit, Gross2009oou,
Ottenstein2008cso, Huckans2009tbr, Wild2012mot}.

The new Efimov resonances used to demonstrate universality in Ref.\
\cite{Berninger2011uot} are in the vicinity of two open-channel-dominated
$s$-wave Feshbach resonances that were theoretically predicted with pole
positions around 550~G and 800~G \cite{Chin2010fri}. Model M2004 is quite
accurate at fields below 100~G, but at higher fields its predictions are in
error by up to $\approx$ 1 per cent, or 8~G for $B$ on the order of 800~G. This
is not accurate enough to describe the scattering physics to sufficient
precision to interpret the Efimov resonances, so an improved theoretical model
is essential.

The aim of the present paper is to describe the previously unexplored
properties of $^{133}$Cs in its absolute atomic ground state in the
magnetic high-field region between 450~G and 1000~G. We present new
measurements of both resonance positions and binding energies at fields
up to 1000~G. We then use these results to develop a new optimized
theoretical model, which we call M2012, that is accurate at both high
and low magnetic fields. This model provides an accurate predictive
tool to map the scattering length $a(B)$ as a function of magnetic
field $B$ \cite{SuppMat}, which it is difficult to measure directly. This mapping was
key to interpreting the three new Efimov features reported in Ref.\
\cite{Berninger2011uot}.

This paper is organized as follows.  Section~\ref{CsThresh} describes the
essential molecular physics of threshold Cs states and describes the notation
we use. Section~\ref{Experiment} describes our experimental methods and results
in the high-field region. Section~\ref{Theory} describes our theoretical model.
Section~\ref{Fit} describes our least-squares fits to the experimental results,
and compares experiment with theory in the regions of high, middle, and low
fields. Section~\ref{Conclusion} summarizes our conclusions.

\section{Near-threshold states of Cesium dimer}
\label{CsThresh}

\begin{figure}[htbp]
\includegraphics[width=\columnwidth]{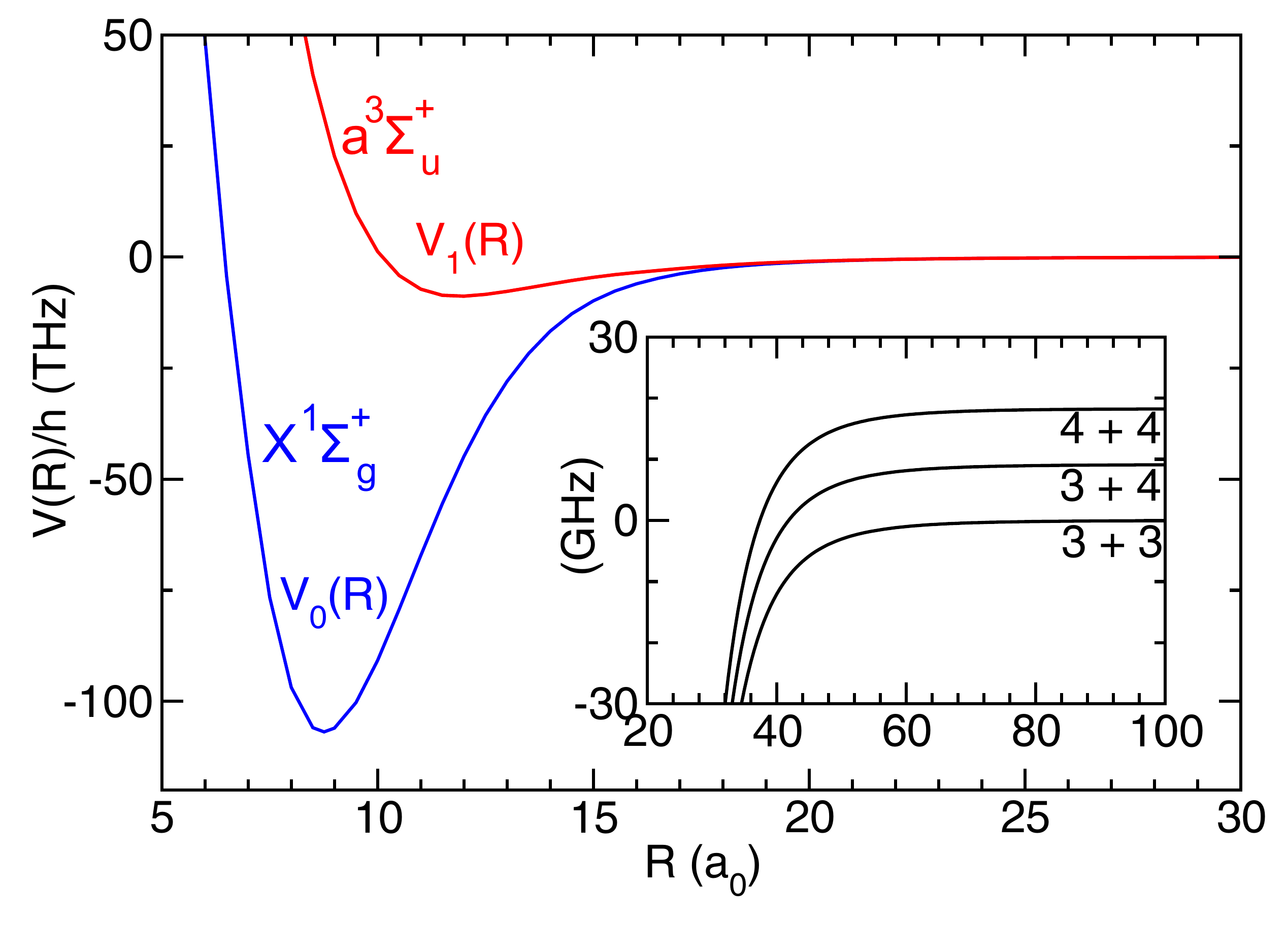}
\caption{(Color online) Molecular potential energy curves $V_0(R)$ and
$V_1(R)$ for the singlet and triplet states of
Cs$_2$. The inset shows an expanded view of the
long-range potentials separating to the two different $f=$ 3
and 4 hyperfine states of the atoms at magnetic field $B=0$.} \label{fig-curves}
\end{figure}

Two alkali-metal atoms in $^2$S states interact at short range to form singlet
($X\,^1\Sigma^+$) and triplet ($a\,^3\Sigma^+$) states, with potential curves
as shown in Figure \ref{fig-curves}. Levels that lie more than about 100 GHz
below dissociation have fairly well-defined singlet or triplet character, so
lie principally on one or the other of these curves. However, the levels of
primary interest in the present work are very close to dissociation, and are
bound by less than 1 GHz (and sometimes as little as 10 kHz). In this region
the singlet and triplet states are strongly mixed by hyperfine interactions and
it is more appropriate to describe the levels in terms of atomic quantum
numbers.

The zero-field levels of the Cs atom are characterized by the nuclear
spin $i=7/2$, the electron spin $s=1/2$ and their resultant $f=3$ or 4,
with the $f=4$ level 9.19 GHz above $f=3$. In a magnetic field $B$,
each level splits into 2$f$+1 sublevels labeled by $m_f$, with the
ground state $(f,m_f)=(3,+3)$. In the present work we label the atomic
states with letters ${\rm a,b,c,}\ldots$ in increasing order of energy.

For Cs$_2$ there are 3 field-free atomic thresholds, labeled in
increasing order of energy by $(f_1,f_2)$ = (3,3), (3,4) and (4,4), as
shown in the inset of Fig.\ \ref{fig-curves}. In a magnetic field, each
threshold splits into sublevels labeled by $(f_1,m_1) + (f_2,m_2)$. The
near-threshold molecular states are to a good approximation described
by quantum numbers $(f_1,f_2,F,M_F)$, where $F$ is the resultant of
$f_1$ and $f_2$ and $M_F=m_1+m_2$ (though $m_1$ and $m_2$ are not
individually conserved). $M_F$ is a nearly good quantum number except
near avoided crossings. For a homonuclear molecule such as Cs$_2$, $F$
is also nearly conserved in the region where the atomic Zeeman effect
is near-linear. Additional quantum numbers are needed for the molecular
vibration $n$ and the partial-wave angular momentum $L$.  For
near-dissociation levels it is convenient to specify $n$ with respect
to dissociation, so that the topmost level is $n=-1$, the next is
$n=-2$, etc. In the present work we describe near-threshold levels
using a set of quantum numbers $n(f_1f_2)FL(M_F)$, with $L=0,2,4,$
etc.\ indicated by the usual labels $s$, $d$, $g$, etc. This is
sometimes abbreviated to $FL(M_F)$ to avoid repetition. Following
Ref.~\cite{Chin2010fri}, we speak of bound levels with dominant $s$
character in their wavefunction as $s$-wave levels; similarly for $d$-
or $g$-wave levels with dominant $L=2$ or 4 character in their
wavefunctions.

Each molecular level lies within a ``bin" below its associated threshold, with
the boundaries of the bins determined by the long-range forces between the
atoms. For Cs$_2$, with $V(r)=-C_6r^{-6}$ at long range and $C_6\approx 6890\
E_{\rm h}a_0^6$ for both the singlet and triplet states, the $n=-1$ level lies
between zero and $-105$ MHz, and the $n=-2$ level lies between $-105$ MHz and
$-725$ MHz~\cite{Chin2010fri}. Similarly, bin boundaries can be worked out for
more deeply bound levels. For Cs$_2$ the background scattering length for each
channel is large and positive, on the order of the scattering length of the
triplet potential. Under these circumstances all the levels lie near the top of
their respective bins, and their energies $E$ are approximately given by those
of the triplet Born-Oppenheimer potential. Numerically, $E/h$ is
$-0.0046$\,GHz, $-0.11$\,GHz, $-0.75$\,GHz, $-2.4$\,GHz, $-5.5$\,GHz,
$-10.6$\,GHz, and $-18.1$\,GHz for $n=-1$ to $-7$ for the M2004 model
\cite{Chin2004pfs}.

\begin{figure*}[htbp]
\includegraphics[width=2\columnwidth,clip]{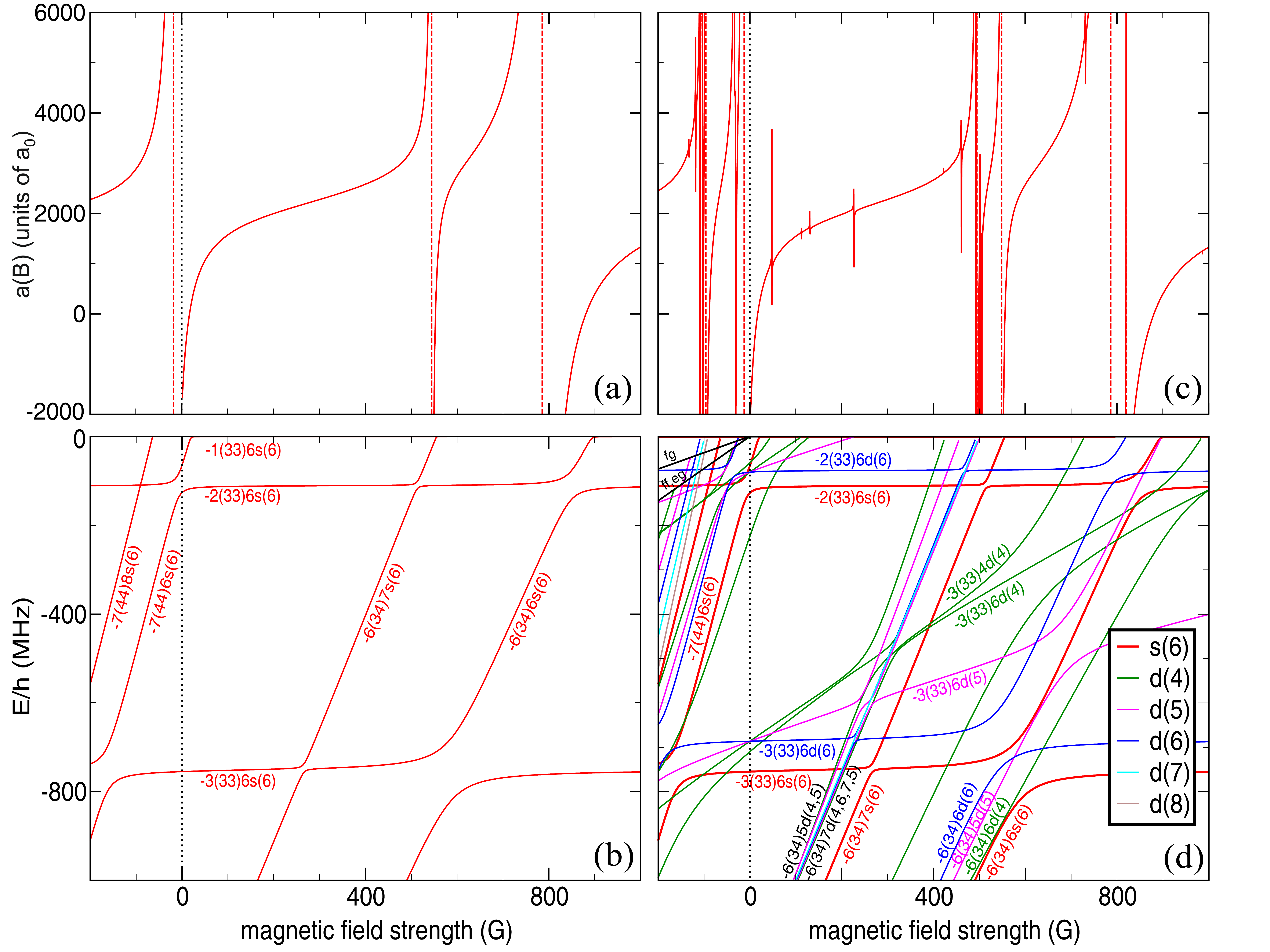}
\caption{(Color online) Scattering length and energy levels versus magnetic field $B$ for Cs collisions in the lowest energy ${\rm aa}$ spin channel. Panel (a): scattering length, calculated including only
$s$ basis functions. The vertical lines indicate the pole positions. Panel (b):
Cs$_2$ $s$-wave bound-state energies below the ${\rm aa}$ threshold with $M_{\rm
tot}=+6$, $M_F=+6$ in the range $B=-200$ to 1000 G, calculated with $s$ basis
functions only. Panel (c): scattering length, calculated including $s$ and $d$
basis functions and including all matrix elements of the spin-dipolar coupling.
Panel (d): Cs$_2$ $s$-wave and $d$-wave bound state energies below the ${\rm aa}$
threshold with $M_{\rm tot}=+6$ and all allowed values of $M_F$ in the range
$B=-200$ to 1000 G. The legend shows the $L(M_F)$ labels. Levels with different
$L$ or $M_F$ cross since small off-diagonal spin-dipolar matrix elements
coupling them were not included in the calculation. Properties in the ${\rm aa}$
channel at negative values of $B$ apply to the ${\rm gg}$ channel with reversed
$M_\mathrm{tot}=-6$ (see text). The lines in the upper left corner of
Panel (d) show the energies of the ${\rm fg}$, ${\rm ff}$, and ${\rm eg}$ atomic channels. }
 \label{figsd6}
\end{figure*}
Feshbach resonances occur where a weakly bound state exists at the same energy
as the colliding atoms. Zero-energy Feshbach resonances thus occur at magnetic
fields where a bound state crosses an atomic threshold. Each resonance is
labeled by the quantum numbers of the bound state that causes it.  We will work
here with the scattering and bound states associated with the ${\rm aa}$ entrance
channel, with two atoms in state ${\rm a}$ with $(f_1,m_1) = (3,+3)$. The energy zero
at any magnetic field strength $B$ is set to the energy of two ${\rm a}$-state atoms.
In $s$-wave scattering, the projection of the total angular momentum onto the
field, $M_{\rm tot}$, is thus always +6 and is a rigorously conserved quantity.
The left-hand panels of Figure~\ref{figsd6} show the $s$-wave bound states
with $M_{\rm tot}=M_F=+6$ at magnetic fields up to 1000 G, together with the
scattering length calculated using only $s$ functions and thus including only
resonances due to bound states with $L=0$.

The $(f_1,f_2)=(3,3)$ levels with $M_F=+6$ in Fig.~\ref{figsd6} have the same
magnetic moment as the separated ${\rm aa}$ atoms, and thus the energies of these
bound states are parallel to the $E=0$ axis. However, $M_F=+6$ levels arising
from other $(f_1,f_2)$ combinations, and (3,3) levels with $M_F\ne+6$, have
different magnetic moments and can cross the ${\rm aa}$ threshold as the magnetic
field is varied. The three strong $s$-wave resonances at fields below 1000 G
are associated with ramping $n(f_1f_2)FL(M_F)$ states of $-7(44)6s(6)$,
$-6(34)7s(6)$ and $-6(34)6s(6)$ character. However, it should be noted that in
each case the ramping $s$-wave state mixes strongly with the least-bound state
$-1(33)6s(6)$, which has a binding energy near 50 kHz, and this mixed state
crosses threshold (and causes a pole in the $s$-wave scattering length) at a
magnetic field below the field where the unperturbed ramping $s$-wave state
would cross threshold. Such shifts in pole position are discussed in
Refs.~\cite{Julienne2004mcm,Julienne2006,Chin2010fri}.

\begin{figure}[htbp]
\includegraphics[width=\columnwidth]{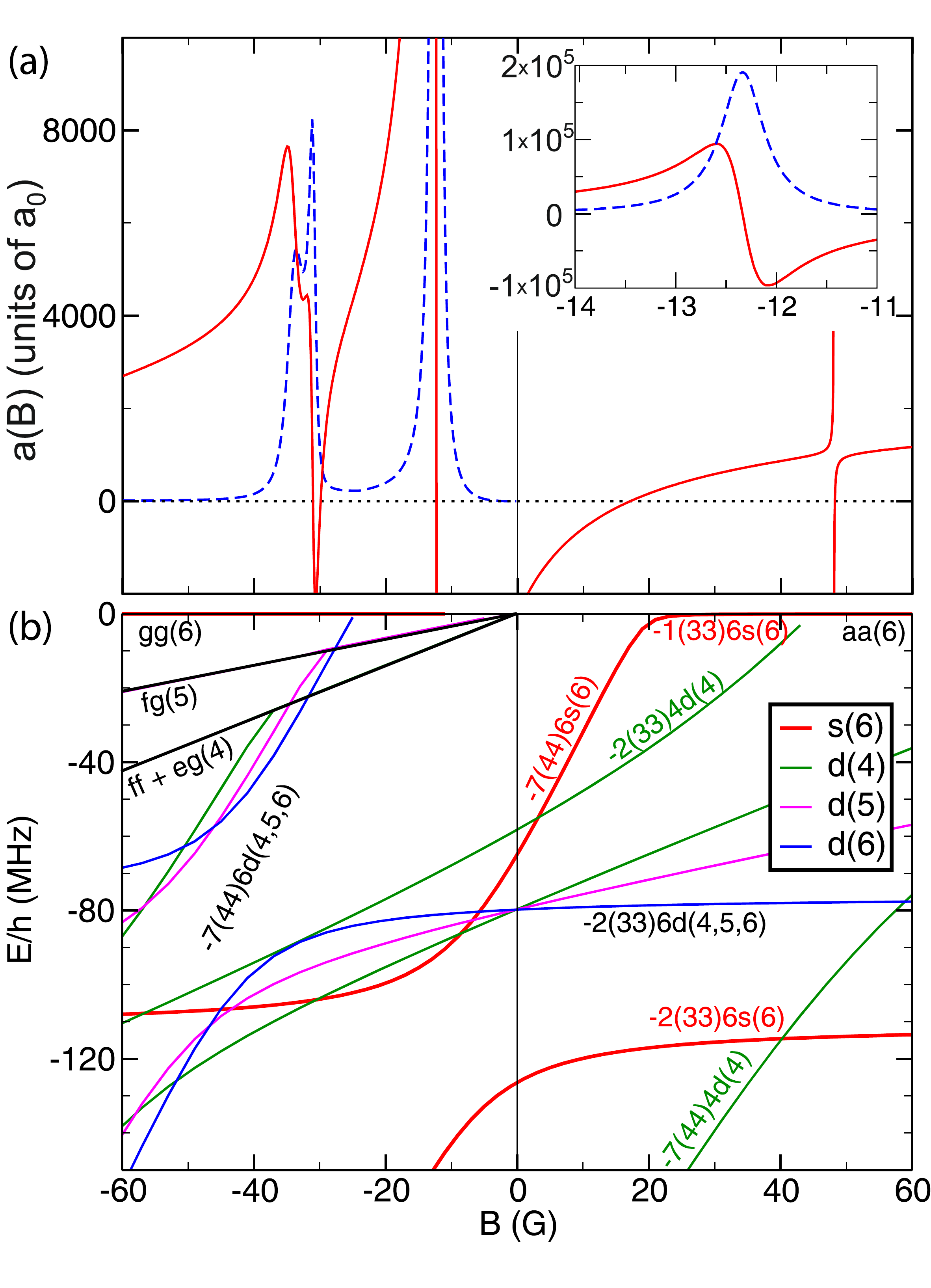}
\caption{(Color online) Panel (a): the real (red, solid) and imaginary (blue,
dashed) parts of the complex scattering length (see Section~\ref{Theory}),
calculated with an $sd$ basis set. For $B<0$, two-body relaxation is possible
to channels ${\rm fg}$, ${\rm ff}$, or ${\rm eg}$. The inset shows an expanded view of the pole
region of the resonance near $-12$~G indicating a maximum variation in
scattering length of $\approx \pm 10^5$ a$_0$ due to resonance decay.  Panel
(b) shows the bound states below threshold, calculated as in Fig.~\ref{figsd6}
so that levels of different $L$ or $M_F$ cross instead of showing avoided
crossings. The $M_F$ labels in the negative $B$ regions are shown with reversed
sign. } \label{figsdlow}
\end{figure}

Figure \ref{figsd6} extends to negative magnetic field. This is to be
interpreted as a reversal of axis, which is equivalent to changing the sign of
all spin projection quantum numbers.  Thus, the ${\rm aa}$ channel at $-|B|$ is
equivalent to the ${\rm gg}$ channel at $+|B|$, where the ${\rm gg}$ channel has two
$g$-state atoms [$(f_1,m_1) = (3,-3)$]. The bound states and scattering length
are continuous across $B=0$, and in particular the low-field behavior of the
scattering length is largely due to the ramping $-7(44)6s(6)$ state, which
actually produces a resonance around $B=-12$ G (i.e., in the ${\rm gg}$ channel), as
shown in detail in Fig.\,\ref{figsdlow}.

\begin{figure}[htbp]
\includegraphics[width=\columnwidth]{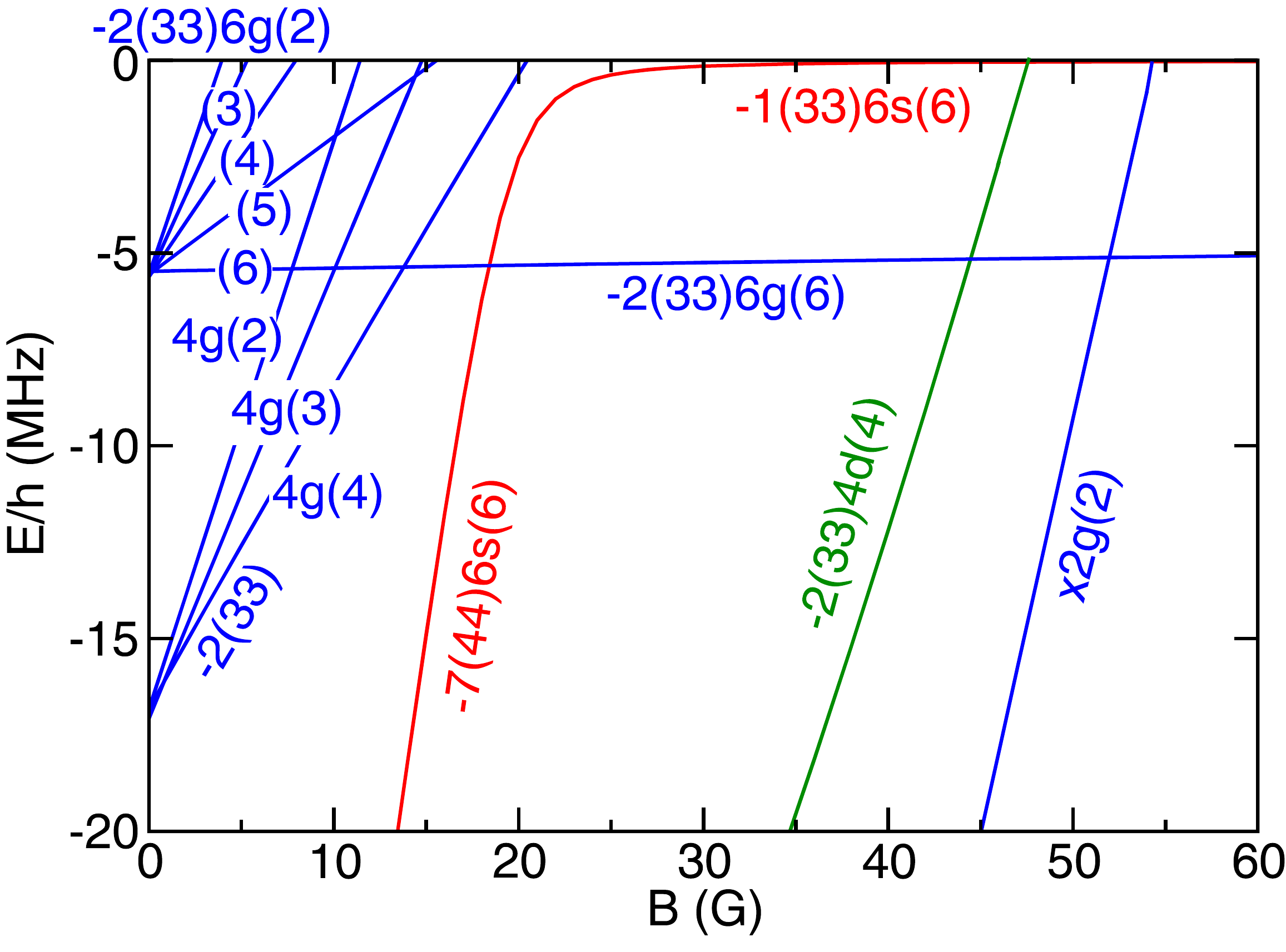}
\caption{(Color online) Low-field near-threshold bound levels of $s$, $d$, and
$g$ symmetry. As for Figs.~\ref{figsd6} and ~\ref{figsdlow}, the avoided
crossings between states of different $L$ or $M_F$ are not calculated because
spin-dipolar coupling is omitted in this calculation. The low-field $g$-wave
levels are of $-2(33)$ character, whereas the level marked $x2g(2)$ is of mixed
$n(f_1f_2)=-2(33)$, $-6(34)$, and $-7(44)$ character. } \label{figlow}
\end{figure}

Each $s$-wave bound state has a corresponding $d$-wave state, also with
$M_F=+6$, that lies almost parallel to it but is shifted by the rotational
energy of the vibrational state concerned; the rotational energy increases with
binding energy and thus depends strongly on the vibrational quantum number $n$.
However, levels with $L>0$ and projection $M_L$ such that $M_{\rm
tot}=M_F+M_L=+6$ will also cross the ${\rm aa}$ threshold and can contribute to
$s$-wave threshold scattering~\cite{Hutson:Cs2-note:2008}. The lower-right panel of
Fig.~\ref{figsd6} shows the bound-state energies including the additional
$d$-wave levels with $M_F \ne 6$. The upper-right panel shows the $s$-wave
scattering length obtained with a basis set including both $s$ and $d$
functions (which we refer to as an $sd$ basis set), showing the additional
resonances that occur. Figure~\ref{figsdlow} shows an expanded view of the
scattering length and near-threshold bound states calculated with an $sd$ basis
set in the low-field region between $-60$ and +60~G. Figure~\ref{figlow} shows
the near-threshold $g$-wave bound states, as studied in
Refs.\,\cite{Chin2004pfs,Mark2007sou,Hutson:Cs2-note:2008}.

Previous work on ultracold Cs has focused on the low-field region. The first
experimental studies of the collisional properties of Cs were performed at
Paris \cite{Soding1998gsr, Gueryodelin1998ibe, Gueryodelin1998seo}, Oxford
\cite{Hopkins2000moe, Arlt1998soc} and Stanford \cite{Vuletic1999ool,
Chin2000hrf, Chin2003sdo}. Chin {\it et al.}~\cite{Chin2000hrf} and Leo {\it et
al.}~\cite{Leo2000cpo} studied over 30 resonances in several spin channels at
fields below 130~G.  Chin {\it et al.}~\cite{Chin2004pfs} observed over 60
resonances in 8 different spin channels (${\rm aa}$, ${\rm gg}$, ${\rm hh}$, ${\rm an}$, ${\rm ao}$, ${\rm ap}$,
${\rm hn}$ and ${\rm gf}$) for fields up to 250~G. In recent years, we have explored the
energy spectrum of weakly bound Cs$_{2}$ Feshbach molecules by magnetic moment,
microwave \cite{Mark2007sou} and magnetic field modulation spectroscopy
\cite{Lange2009doa}. The zero crossing of the $s$-wave scattering length was
also precisely determined, using an approach based on measuring the
interaction-induced dephasing of Bloch oscillations \cite{Gustavsson2008coi,
Gustavsson2008PhD}.


\section{Feshbach spectroscopy at high magnetic field}\label{Experiment}

In the present work, we carry out a variety of different experiments on
ultracold $^{133}$Cs in its lowest internal state, ($f,m_f$)=(3,3), at magnetic
fields in the range between 450 and 1000\,G. We first discuss the main
experimental procedures and conditions (Sec.\,\ref{Sec:SamplePreparation}).
Next we report on trap-loss spectroscopy, which allows measurement of the
positions of narrow Feshbach resonances (Sec.\,\ref{Sec:FeshbachSpectroscopy}).
Finally we present magnetic field modulation spectroscopy for the precise
determination of molecular binding energies near broad Feshbach resonances
(Sec.\,\ref{Sec:BindingEnergyMeasurements}).

\subsection{Sample preparation} \label{Sec:SamplePreparation}
To access the high-field region, we have implemented a new magnetic-field
system in the experimental setup \cite{Berninger2011PhD}. This system is able
to reach maximum magnetic field strengths up to 1400\,G in a steady-state
condition with 10\,mG long-term stability. The high magnetic bias fields are
created by three separately controllable pairs of magnetic field coils, made of
4\,mm and 6\,mm square-profile copper tubes insulated by glass-fiber braided
sleevings. For each coil pair the electric currents, which are up to 400\,A
(4\,mm tube) and 800\,A (6\,mm tube), respectively, are supplied by two 6\,kW
power supplies, connected in parallel. The temperature of the coil system is
kept below 50$^{\circ}$C by internal water cooling of the copper tubes using a
10-bar pump system. Magnetic field stability is governed by controlling the
current in the coils by an active feedback system, which operates at a
precision level of $10^{-5}$. For this, the actual currents are measured by
highly sensitive current transducers. We have checked that precise current
control is sufficient for magnetic field control to the limit given above.
Other influences, such as thermal expansion of the copper coils, play a minor
role. A detailed description of the magnetic-field coil system can be found in
Ref.\,\cite{Berninger2011PhD}.

The procedure used to prepare an ultracold cesium sample in the absolute atomic
ground is based on well established cooling and trapping techniques, which are
similar to the ones described in Ref.\,\cite{Kraemer2004opo} down to the
$\mu$K-regime. After Zeeman slowing and cooling in a magneto-optical trap, the
atoms are loaded into a three-dimensional optical lattice created by four laser
beams, where Raman-sideband cooling \cite{Weber2003PhD, Kerman2000bom,
Treutlein2001hba} is performed for 3.5\,ms. During this stage, where a small
magnetic field of several hundred mG is applied, the atoms are cooled and
spin-polarized into the absolute ground state. After Raman-sideband cooling the
ensemble size amounts to $1.5 \times 10^{7}$ atoms at a temperature of about
$1\,\mu$K. Then, the atoms are transferred into a large-volume far-off-resonant
dipole trap \cite{Grimm2000odt} generated by two crossed 100\,W CO$_{2}$ laser
beams, featuring a waist of about 600\,$\mu$m each. As the optical trap is not
strong enough to hold the atoms against gravity, an additional magnetic
levitation field of 31\,G/cm is applied~\cite{Weber2003bec}.

\begin{figure}
\includegraphics[width= \columnwidth]{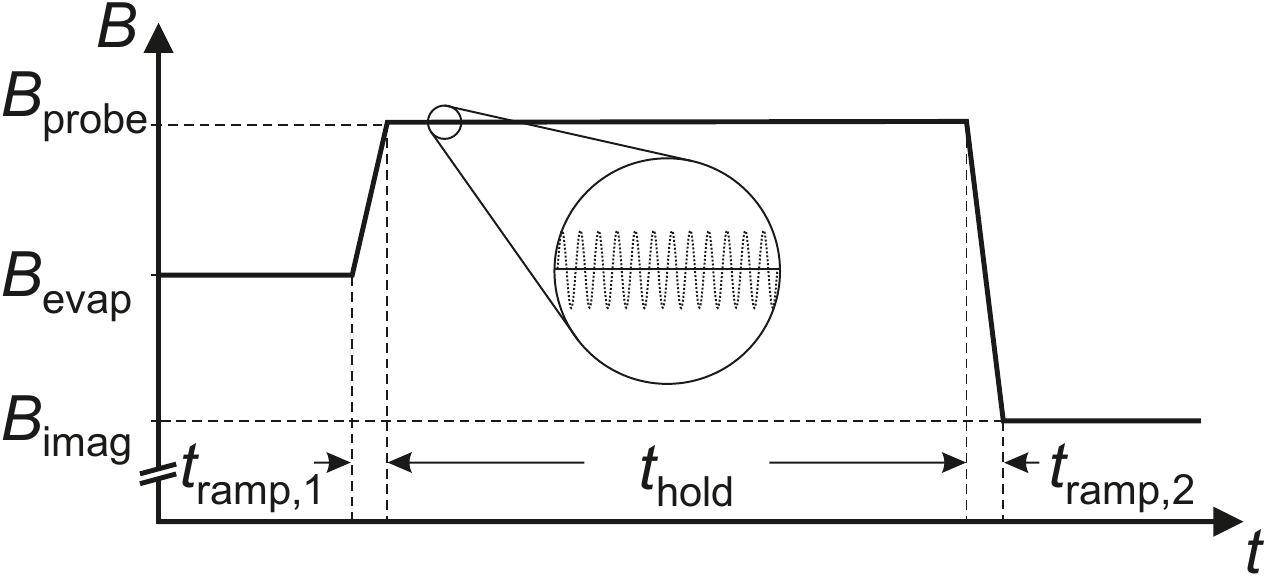}
\caption{Timing sequence for the magnetic field ramps for trap-loss
spectroscopy and binding energy measurements. The magnetic field strength $B$
is linearly ramped from the final evaporation field $B_{\rm evap}$ within the
ramping time $t_{\rm ramp,1}$ to the probe field $B_{\rm probe}$. After an
experimentally optimized hold time $t_{\rm hold}$ ranging between 0.2 and 1\,s,
the magnetic field strength is (linearly) ramped to the imaging field $B_{\rm
imag}$. For {\it Scheme~A}, $B_{\rm evap}$ and $B_{\rm imag}$ are in the
magnetic low-field region. The magnified segment visualizes $B_{\rm probe}$,
which is constant in time for trap-loss spectroscopy (solid line) and
sinusoidally modulated for the determination of binding energies (dashed
line).} \label{fig:Ramping_scheme}
\end{figure}

We use two different schemes, which we refer to as {\it Scheme~A} and {\it
Scheme~B}. {\it Scheme A} is based on evaporation and detection at low magnetic
fields, similar to our previous work \cite{Mark2007sou}. This scheme could be
implemented in a simple way, but ramping up to the high probe field and back
down to the detection field involves crossing several Feshbach resonances,
which causes additional losses and heating. In the course of performing the
present experiments, we developed an improved approach ({\it Scheme B}) that
allows imaging in the magnetic high-field region and optimization of
evaporative cooling at higher fields. In the following, both schemes are
described in detail. Figure \ref{fig:Ramping_scheme} illustrates the generic
timing sequence for both schemes.

In {\it Scheme A}, evaporation is performed for 2\,s at constant depth of the
CO$_2$ laser trap in the magnetic low-field region. This stage of plain
evaporation results in $\approx 5 \times 10^{6}$ atoms at a temperature
slightly below $1\,\mu$K. Then, the CO$_2$ laser trap is spatially overlapped
with a crossed dipole trap created by a 1064\,nm fiber laser, with waists of
$40\,\mu$m and $250\,\mu$m. To continue evaporation, the tightly focussed
$40\,\mu$m beam is ramped down from 60\,mW to 3.5\,mW within 6.5\,s, while the
intensity of the $250\,\mu$m beam is fixed at 400\,mW. During this procedure,
both CO$_2$ laser beams are switched off, finishing the sample transfer. The
$s$-wave scattering length is large and positive during the evaporation
sequence and is adjusted for the final evaporation step to $a \approx
200\,a_0$, corresponding to a magnetic field strength of $B_{\rm evap} \simeq
21$\,G. Efficient evaporation conditions are encountered at this field because
of an Efimov-related three-body recombination loss minimum
\cite{Kraemer2006efe}. In this way, we end up with $10^5$ thermal atoms at a
temperature of 70\,nK in the magnetic low-field region.

When ramping over the broad Feshbach resonances, sizeable effects of three-body recombination are unavoidable, even when applying the fastest possible ramp speeds. This causes direct recombination losses and additional heating \cite{Weber2003tbr}, which can cause subsequent evaporation losses in the measurement process. To avoid the latter effect, we recompress the trap by increasing the intensity
of the $40\,\mu$m beam by about a factor 10 before the magnetic field ramp is
carried out. In the final step of the sample preparation, the
levitation field is decreased to 8\,G/cm.
The mean trap frequency is about $\bar{\omega}=2\pi \times 46(5)$~Hz, and the
final sample contains $10^5$ thermal atoms at a temperature of 120\,nK in the
magnetic low-field region.

The measurements are performed by linearly ramping from $B_{\rm evap}$ within a
ramp time $t_{\rm ramp,1}=10$\,ms to the probe fields $B_{\rm probe}$ in the
magnetic high-field region. As described above, crossing of the broad $s$-wave
Feshbach resonances leads to considerable heating of the sample and additional
particle loss. We estimate the temperature at $B_{\rm probe}$ to be between
150\,nK and 200\,nK with {\em Scheme A}.

To determine the particle number, we linearly decrease the magnetic bias field
to zero ($B_{\rm imag}\simeq 0$) within $t_{\rm ramp,2}=10$\,ms and carry out
resonant absorption imaging. The temperature is obtained in time-of-flight
expansion measurements after release from the trap. The magnetic field strength
is determined from measurements of the $(3,3) \rightarrow (4,4)$ microwave
transition frequency by applying the Breit-Rabi formula \cite{Breit1931mon}.

In {\it Scheme~B}, after loading the atoms from the Raman lattice to the
levitated CO$_2$ laser trap, the magnetic field is linearly ramped to the magnetic
high-field region within 10\,ms. For measurements performed below 800\,G the
ramp ends at 561\,G ($a=1090\,a_0$), whereas for measurements above 800\,G it
ends at 970\,G ($a=1140\,a_0$).

At this stage, the three-body recombination losses that are encountered while
crossing the broad $s$-wave Feshbach resonances are limited because of the low
density and the relatively high temperature ($T \simeq 1\,\mu$K) of the sample.
The temperature dependence of losses follows from the unitarity limitation of
three-body recombination rates \cite{Dincao2004lou}. To compensate for the
small change in the magnetic moment that is encountered during the ramp as a
consequence of the quadratic contribution to the Zeeman effect, the magnetic
levitation field is adjusted simultaneously. Then,  2~s of plain evaporation
result in $\approx 5 \times 10^{6}$ atoms at a temperature of about $1\,\mu$K
(similar to {\it Scheme~A}). Then, the CO$_2$ laser trap is spatially
overlapped with the 1064\,nm crossed dipole trap as described in {\it
Scheme~A}.

In the crossed dipole trap, we perform forced evaporation by decreasing the
laser intensity of the $40\,\mu$m beam from 60\,mW to 3.5\,mW within 15\,s.
During this step, the intensity of the $250\,\mu$m beam is changed only
slightly, from 400\,mW to 300\,mW. Both CO$_2$ lasers are switched off during
the first 5\,s of evaporation to achieve an efficient transfer to the 1064\,nm
trap. The magnetic bias field is adjusted during the evaporation sequence for
optimized elastic scattering conditions. The last evaporation step of this
sequence ends at $B_{\rm evap}=558.7$\,G ($a \approx 700\,a_0$) or $B_{\rm
evap}=894$\,G ($a \approx 300\,a_0$), respectively. Note that close to 894\,G
an Efimov-related recombination minimum is present~\cite{Berninger2011uot},
which apparently facilitates efficient evaporation. After recompression and
reshaping, leading to a mean trap frequency of $\bar{\omega}=2 \pi \times
26(3)$\,Hz, we end up with a non-condensed sample of between $5 \times 10^4$
and $10^5$ atoms at a temperature of about 50\,nK.

For {\it Scheme~B}, no broad $s$-wave Feshbach resonances are crossed in the
final magnetic field ramps to reach $B_{\rm probe}$, and therefore no
noteworthy heating effects and particle losses are observed. The particle
number is determined by high-field imaging in the vicinity of the
zero crossing of the broad $s$-wave Feshbach resonances at $B_{\rm
imag}=556.4$\,G and $B_{\rm imag}=887.5$\,G. The magnetic field ramps to and
from $B_{\rm probe}$ involve linear changes of the magnetic field with $t_{\rm
ramp,1}=t_{\rm ramp,2}=10$\,ms, as for {\it Scheme A}. The procedures for
magnetic field calibration and temperature determination are the same as the
ones in {\it Scheme~A}.

\subsection{Trap-loss spectroscopy}\label{Sec:FeshbachSpectroscopy}
Trap-loss spectroscopy is a well established method based on the enhancement of
collisional losses in the vicinity of a Feshbach resonance \cite{Chin2010fri}.
For atoms in the absolute atomic ground state, where inelastic binary collisions
are energetically forbidden, three-body recombination is the lowest-order loss
process. In this process three colliding atoms recombine to a molecule and a
free atom. Typically, the kinetic energy released far exceeds the trapping
potential, leading to loss of the three particles involved. The general
$a^4$-scaling of three-body recombination rates \cite{Fedichev1996tbr,
Esry1999rot, Nielsen1999ler, Weber2003tbr} leads to a maximum in losses at the
magnetic field position $B_{\rm max}$, corresponding to the divergence of $a$
at the Feshbach resonance pole, and a minimum at the position $B_{\rm min}$,
close to the zero crossing of the $s$-wave scattering length. This allows the
observed losses to be directly related to the positions and widths of the
Feshbach resonances.

\begin{figure}
\includegraphics[width= \columnwidth]{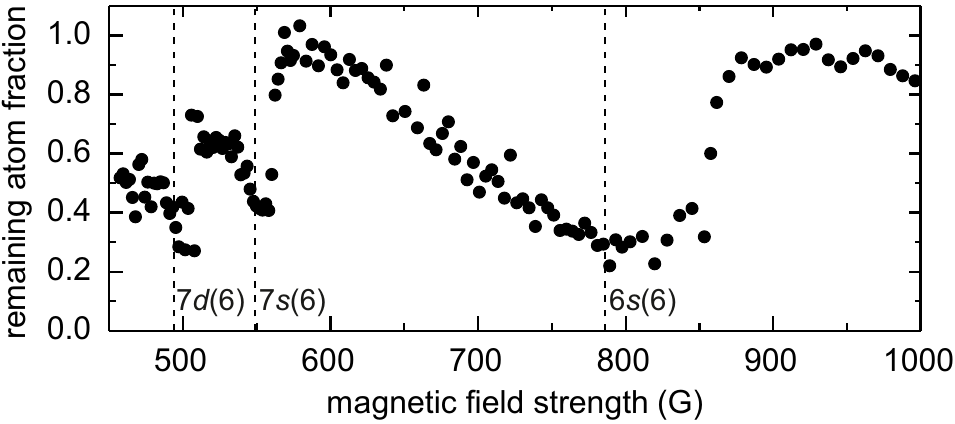}
\caption{Results of trap-loss spectroscopy in the magnetic field region between
450\,G and 1000\,G, performed with {\it Scheme~B}. The enhancements in
losses result from broad Feshbach resonances centered at 494\,G, 549\,G and
787\,G. The resonance positions, which are derived from our theoretical model
(see Sec.\,\ref{Theory}), are indicated by dashed lines, labeled according to
the quantum numbers $F L (M_{F})$ of the molecular states that cause the
resonances. The measurements are performed with a hold time of 500\,ms. Narrow
Feshbach resonances are not visible in this scan because of the large step size
of about 2\,G. A remaining atom fraction of $1.0$ corresponds to $8 \times 10
^{4}$~atoms.} \label{fig:FeshbachScanSwave}
\end{figure}

\begin{figure*}
\includegraphics[width= 2\columnwidth]{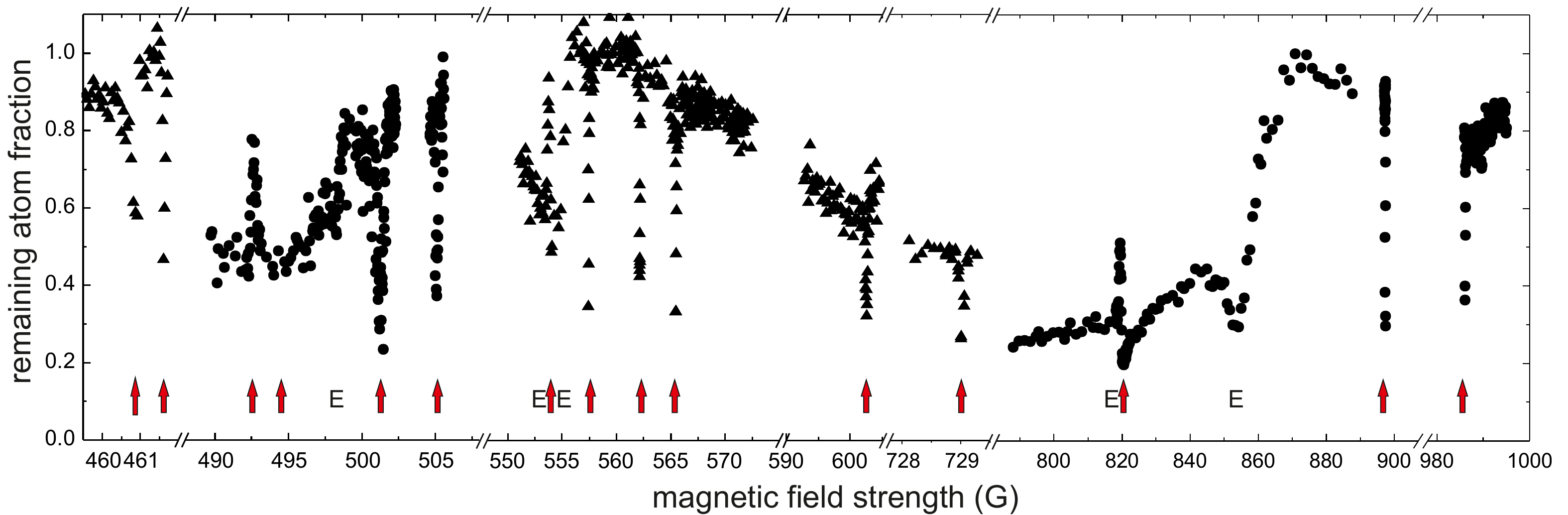}
\caption{(Color online) Detailed results of trap-loss spectroscopy in the
magnetic high-field region. We observe 15 Feshbach resonances, stemming from
$d$-, $g$- and $i$-wave molecular states. Measurements indicated by
$(\blacktriangle)$-symbols are obtained by {\it Scheme~A}, whereas the
$(\bullet)$-symbols refer to data points acquired with {\it Scheme~B}. The
poles of the Feshbach resonances, omitting the $s$-wave resonances, are marked
with an arrow. The loss features at 498.1\,G, 553.3\,G, 554.7\,G, 818.9\,G and
853.1\,G, which are indicated by ``E'', are related to Efimov loss resonances
as reported in Refs.\,\cite{Berninger2011uot, Ferlaino2011eri}. Note that the
data in the intervals [728,729.5]\,G and [980,1000]\,G, which are measured for
different $t_{\rm hold}$, are multiplied by scaling factors of 0.5 and 0.9,
respectively, to reproduce the overall behavior shown in
Fig.\,\ref{fig:FeshbachScanSwave}. A remaining atom fraction of $1.0$
corresponds to $8 \times 10 ^{4}$~atoms.} \label{fig:FeshbachScan}
\end{figure*}

In this Section, we will first report on experiments characterizing the
scattering properties in the vicinity of the broad $s$-wave Feshbach
resonances, as shown in Fig.\,\ref{figsd6}(a), by performing a broad magnetic
field scan with large step size. Then, we decrease the step size to perform
detailed scans to identify and characterize narrow Feshbach resonances, which
originate from states with higher rotational angular momentum ($L > 0$).

Trap-loss spectroscopy is performed by recording the remaining atom fraction
after a hold time $t_{\rm hold}$ at the probe field $B_{\rm probe}$. In
general, we cannot exclude additional losses encountered within $t_{\rm
ramp,1}$ and $t_{\rm ramp,2}$ during the magnetic field ramps $B_{\rm evap}
\rightarrow B_{\rm probe}$ and $B_{\rm probe} \rightarrow B_{\rm imag}$.
However, for the characterization of narrow Feshbach resonances only a small
magnetic field region is investigated, where variations in the initial atom
number are negligible. Furthermore, the measurements are performed with $t_{\rm
hold} \gg t_{\rm ramp,1}, t_{\rm ramp,2}$, strongly limiting the effect of
finite ramp times.

The broad scan of the magnetic high-field region covers a range from 450 to
1000\,G, as shown in Fig.\,\ref{fig:FeshbachScanSwave}. This scan clearly shows
two broad loss features around 550\,G and 800\,G, which can be assigned to the
two high-field $s$-wave Feshbach resonances, as discussed in
Sec.\,\ref{CsThresh} (see upper panel Fig.\,\ref{figsd6}). These measurements
demonstrate the large width of the Feshbach resonance near 800\,G. Because of
the unitarity limitation of three-body recombination losses
\cite{Dincao2004lou}, it is not possible to determine $B_{\rm max}$ accurately
for the $s$-wave resonances by trap-loss spectroscopy.

In the region around 500\,G, no $s$-wave Feshbach resonance is expected, but
the theoretical model predicts a series of closely adjoining $d$-wave Feshbach
resonances, as shown in Fig.\,\ref{figsd6}. One of these has a width of about
5\,G, producing the broad loss signal around 495\,G seen in
Figs.\,\ref{fig:FeshbachScanSwave} and \ref{fig:FeshbachScan}.

We perform high-resolution scans by decreasing the step size of the magnetic
field scans to a few mG. The results of these scans are displayed in
Fig.\,\ref{fig:FeshbachScan}. We observe 15 narrow loss features, which can be
assigned according to the theoretical model given in Sec.\,\ref{Theory} to
Feshbach resonances originating from the coupling of the free atoms to
molecular states with rotational angular momentum $L > 0$.

We observe three narrow resonances that cannot be attributed to $s$-, $d$- or
$g$-wave molecular states in the present model. They are found at 461.62\,G,
557.45\,G and 562.17\,G. Our model, however, predicts the existence of Feshbach
resonances stemming from $i$-wave molecular states ($L = 6$) in the magnetic
field regions where we observe these features. The calculations are not
accurate enough to establish an unambiguous assignment, but the match between
experiment and theory nevertheless provides strong evidence that this is the
first experimental observation of $i$-wave Feshbach resonances. These
resonances are discussed further in Sec.\,\ref{Theory}.

The positions of the poles of the $d$-, $g$- and $i$-wave Feshbach resonances
obtained in these measurements are summarized in Table\,\ref{tab:FeshbachScan}.
The peak positions are determined by Gaussian fits to the loss peaks. For
several of these resonances we also identify recombination loss minima, which also provide estimates for the corresponding resonance widths.

\begin{table}
\begin{center}
\begin{tabular}{|c | c | c|}
\hline
molecular state& \multirow{2}{*}{B$_{\rm max}$(G)}  & \multirow{2}{*}{B$_{\rm min}$(G)} \\
$n (f_1 f_2) F L (M_F)$ & &\\
\hline
-6(34)5$d$(5) & 460.86(5) & \\			
$i$-wave$^{1)}$  & 461.62(5) & \\
-6(34)7$d(x)^{2)}$ & 492.45(3) & 492.63(3) \\
-6(34)7$d(x)^{2)}$  & 494.4(9)  & 499.4(1) \\
-6(34)7$d(x)^{2)}$  & 501.24(3) & \\
-6(34)7$d(x)^{2)}$  & 505.07(3) & \\			
$g(3)^{3)}$  & 554.06(2) & 553.73(2) \\
$i$-wave$^{1)}$   & 557.45(3) & \\
$i$-wave$^{1)}$ & 562.17(3) & \\			
$g(4)^{3)}$  & 565.48(3) & \\
-2(33)6$g$(6)$^{4)}$     & 602.54(3) & \\			
$g(5)^{3)}$ & 729.03(3) & \\			
-6(34)6$d$(6)  & 820.37(20) &	819.41(2) \\
-6(34)5$d$(5) & 897.33(3) & \\			
-6(34)6$d$(4)  & 986.08(3) & \\			
\hline
\end{tabular}
\end{center}
\caption{Results of trap-loss spectroscopy in the magnetic high-field region.
The table shows the magnetic field values for loss maxima (B$_{\rm max}$),
resulting from the poles of the Feshbach resonances, and minima (B$_{\rm
min}$), which are related to the zero crossings of the $s$-wave scattering
length. The numbers in brackets are the experimental $1\sigma$ uncertainties, including
statistical and systematic errors. The assignments of the Feshbach resonances
identify the molecular states that cause the resonances. In the case of very
narrow resonances, the zero crossings could not be determined experimentally.\\
$^{1)}$For the $i$-wave resonances only the $L$ quantum number of the molecular
state is known.\\
$^{2)}$These Feshbach resonances arise from $d$-wave molecular states with
$M_F=4, 5, 6$ and $7$, which are strongly mixed at the atomic threshold.
Therefore, we cannot give simple $M_F$
quantum numbers and use ``$x$'' to indicate the strong coupling.\\
$^{3)}$For these states only the quantum numbers $L$ and $M_F$ are known.\\
$^{4)}$This molecular state is strongly mixed with the state $-6 (3 4) 6 g (6)$.}
\label{tab:FeshbachScan}
\end{table}

\begin{figure}
\includegraphics[width= 1\columnwidth]{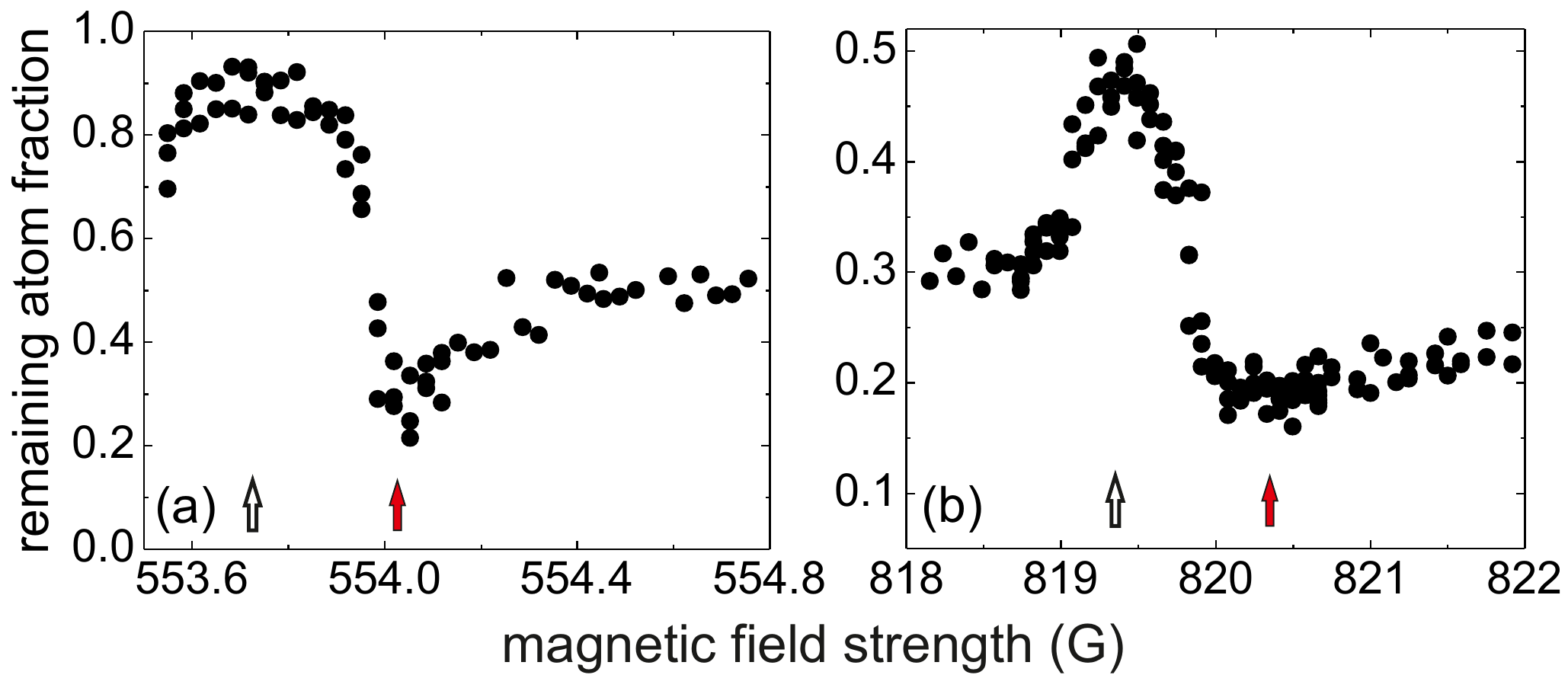}
\caption{(Color online) Expanded view of Fig.\,\ref{fig:FeshbachScan}, showing
the Feshbach resonances arising from the $g$(3) state (a) and the $-6(34)6d(6)$
state (b), which overlap with the broad $s$-wave resonances. The filled and
empty arrows mark the poles and zero crossings of the scattering length,
respectively. The remaining atom fraction at the zero crossing in (b) is
limited due to losses during the magnetic field ramps.}
\label{fig:FeshbachScanBlowUp}
\end{figure}

Figure\,\ref{fig:FeshbachScanBlowUp} shows expanded views of two regions in
Fig.\,\ref{fig:FeshbachScan}, where interesting cases of overlapping FR
scenarios occur. Figure\,\ref{fig:FeshbachScanBlowUp}(a) shows a $g$-wave
resonance centered at 554.06(2)\,G sitting on the shoulder of the 549\,G
$s$-wave Feshbach resonance at a background scattering length of about
$-1000\,a_0$. The zero crossing of $a(B)$ leads to a loss minimum at
553.73(2)\,G. Even more intriguing is a $d$-wave Feshbach resonance situated at
820.4(2)\,G. There, the broad 787\,G $s$-wave Feshbach resonance leads to an
extremely large background scattering length of about $-4200\,a_0$.
Experimentally, this large background masks the loss maximum but clearly
reveals the zero crossing at 819.41(2)\,G, as shown in
Fig.\,\ref{fig:FeshbachScanBlowUp}(b). Both the $g$-wave and $d$-wave resonance
have rather large widths, of 0.33(3)\,G and 0.96(22)\,G, respectively.
Efimov-related three-body physics has been revealed in the vicinity of these
resonances, as reported in Refs.\,\cite{Berninger2011uot, Ferlaino2011eri}.

\subsection{Binding energy measurements}\label{Sec:BindingEnergyMeasurements}
Binding energy measurements of weakly bound dimer states provide a powerful
additional tool to extract information on the cesium interaction potentials and
scattering properties. In particular, for the $s$-wave Feshbach resonances the
exact positions of the poles are obscured by strong loss across a broad
magnetic field range, but can be extracted accurately from binding energies. In
the present work, we measure the binding energies by magnetic field modulation
spectroscopy, a method which was introduced in Ref.~\cite{Thompson2005ump}.
This method is based on a sinusoidal modulation of the magnetic bias field and
allows the creation of dimers starting from an ultracold atom sample. This
leads to an observable loss signal due to fast atom-dimer relaxation when the
modulation frequency matches the binding energy of the dimers plus the small
relative kinetic energy of the colliding atoms. Since the modulation of the
magnetic field is parallel to the magnetic bias field, only transitions between
states with the same projection quantum number of the total angular momentum
are observed. This procedure has been successfully applied in several
experiments \cite{Thompson2005ump, Weber2008aou, Lange2009doa,
Thalhammer2009cam, Beaufils2010rfa, Pasquiou2010cod, Gross2010nsi} to determine
atomic scattering properties.

The atom samples for the binding energy measurements are prepared according to
{\it Scheme~B}, as described in Sec.\,\ref{Sec:SamplePreparation}. At $B_{\rm
probe}$, a modulation signal is applied for a variable duration of $t_{\rm
hold}=0.1$ to 1\,s, in a frequency range of 50 to 1600\,kHz and an amplitude
between 0.5 and 3\,G (see Fig.\,\ref{fig:Ramping_scheme}). The amplitude and
the duration of the pulse are experimentally adjusted for each binding energy
measurement to optimize the signal-to-noise ratio. The signal is generated by a
programmable frequency generator and subsequently amplified by a commercial
25\,W radio-frequency amplifier, which drives the current in a separate set of
coils and thereby creates the modulation of the magnetic bias field.

\begin{figure}
\includegraphics[width= \columnwidth]{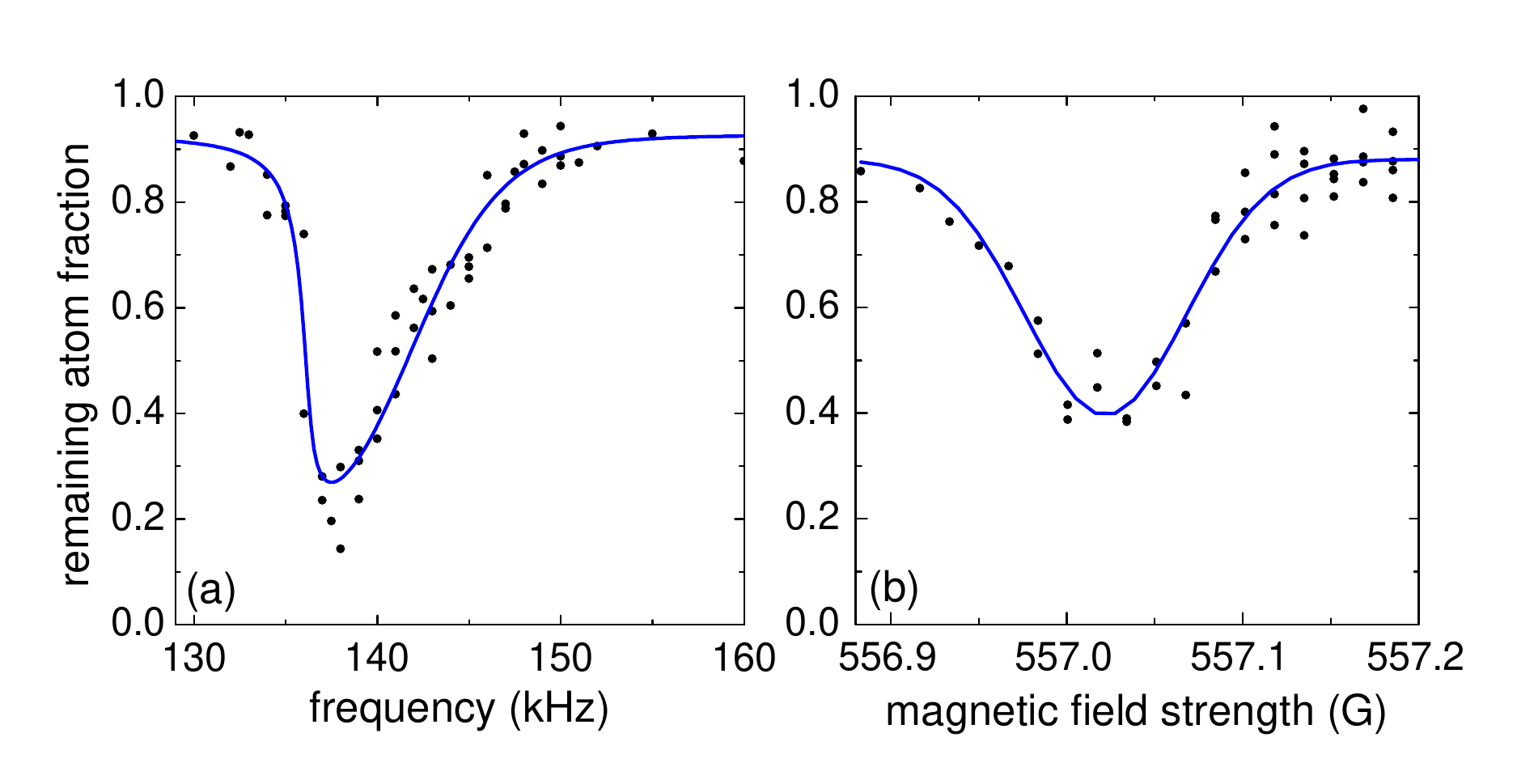}
\caption{(Color online) Typical signals for magnetic field modulation
spectroscopy. (a) Frequency scan at a fixed magnetic field strength of
911.69\,G. The asymmetric shape of the signal is fitted by a model that takes
the temperature of the sample into account \cite{Napolitano1994lso}. (b)
Magnetic field scan at a constant modulation frequency of 425\,kHz, revealing a
loss resonance at 557.02\,G. This resonance stems from the $s$-wave molecular
state. The line represents a Gaussian fit. A remaining atom fraction of 1.0
corresponds to (a) $4.5 \times 10^{4}$ and (b) $2.5 \times 10^{4}$ atoms.}
\label{fig:WiggleSample}
\end{figure}

The measurements are usually performed by varying the modulation frequency at a
fixed $B_{\rm probe}$. Another possibility, however, is to scan $B_{\rm probe}$
while the modulation frequency is kept constant. The advantage of the latter
approach is that it is less sensitive to atom losses caused by technical
imperfections, such as resonance phenomena in the electric circuit that drives
the transitions. We checked that the two methods give consistent results in our
measurements. Figure\,\ref{fig:WiggleSample} shows sample loss signals derived
in a frequency scan (a) and a magnetic field scan (b).

\begin{figure}
\includegraphics[width= \columnwidth]{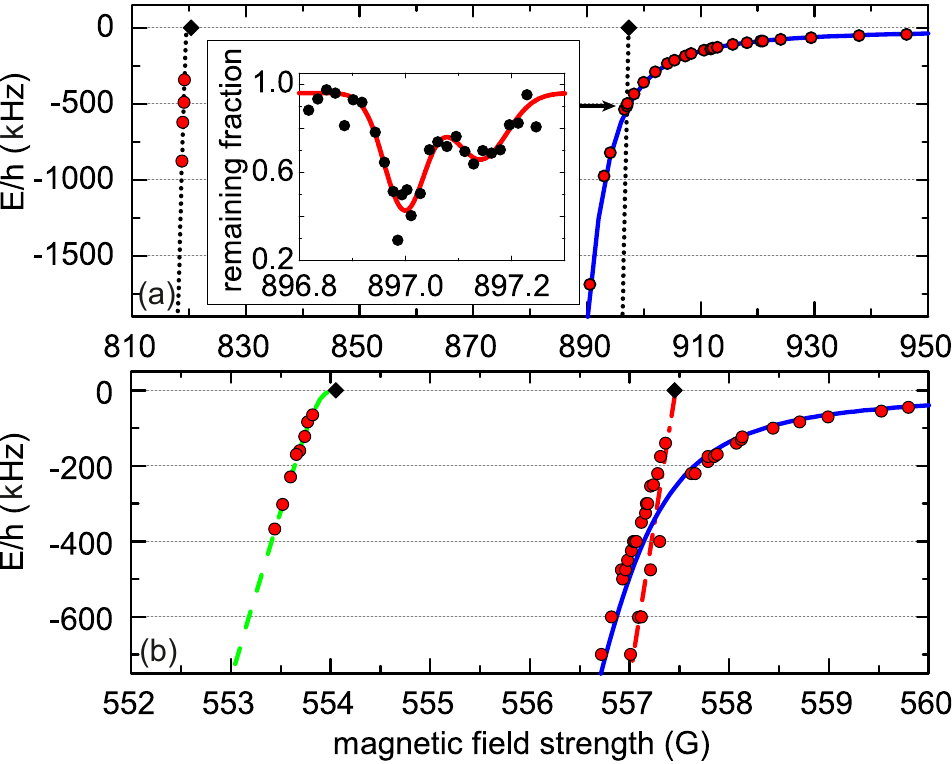}

\caption{(Color online) Results of experimental binding energy measurements.
(a) the $-6(34)6d(6)$ and $-6(34)5d(5)$ states (dashed lines) and the
$-6(34)6s(6)$ state (solid line) between 820\,G and 950\,G. The lines are
guides for the eye. The inset shows a binding energy measurement at the
crossing of the $-6(34)6s(6)$ and $-6(34)5d(5)$ states at 509\,kHz. At fields
above 900\,G the $s$-wave state takes on the character of the least-bound state
$-1(33)6s(6)$, which has a binding energy around $h\times 50$\,kHz away from
avoided crossings. The strong variation of the magnetic moment follows from the
avoided crossing with this threshold $s$-wave state. (b) the binding energies
of the $g$(3) state (dashed line), an $i$-wave state (dot-dash line) and the
$-6(34)7s(6)$ state (solid line). The positions of the resonance poles are
marked by a ($\blacklozenge$)-symbol, with values taken from
Table\,\ref{tab:FeshbachScan}. } \label{fig:BindingEnergies550}
\end{figure}

We studied the binding energies $E_{\rm b}$ of the high-field $s$-wave states
and of several $d$-, $g$- and $i$-wave states, as shown in
Fig.\,\ref{fig:BindingEnergies550}. For the $s$-wave states with $E_{\rm b} / h
< 200$\,kHz, we observed asymmetric line shapes resulting from the finite
temperature of the samples. We include this effect in our fitting routine using
the line-shape model of Ref.\,\cite{Napolitano1994lso}. For $s$-wave states
with $E_{\rm b} / h > 200$\,kHz and for dimer states with higher rotational
angular momentum, the binding energy has a strong dependence on the magnetic
field. In these cases, the magnetic field noise and the field gradient that is
applied to levitate the atoms broaden and symmetrize the loss signals. For
these symmetrized signals, the effect from the finite temperature plays a minor
role, and we therefore obtain $E_{\rm b}$ by fitting a simple Gaussian
distribution to the data.

In the binding energy measurements, we observe several avoided crossings
between molecular states. Around 897\,G and $E_{\rm b} / h \approx 500$\,kHz,
the $-6(34)6s(6)$ state crosses the $-6(34)5d(5)$ state. These two states are
clearly resolved as separate loss features in each magnetic field scan
performed at fixed frequency in the crossing region, as shown in the inset of
Fig.\,\ref{fig:BindingEnergies550}(a). In addition, we observe an avoided
crossing at about 557\,G and $E_{\rm b}/h \approx  350$\,kHz between the
$-6(34)7s(6)$ state and an $i$-wave state, which clearly shows up in the
binding energy measurements presented in Fig.\,\ref{fig:BindingEnergies550}(b).

In view the large difference in the partial-wave angular momentum $\Delta L =
6$, the crossing around 557~G appears to be surprisingly strongly avoided. To
confirm this, we prepare molecular samples in the $s$-wave state by Feshbach
ramps \cite{Chin2010fri} and perform magnetic moment spectroscopy for magnetic
field strengths ranging from 556.5\,G to 557.5\,G using the Stern-Gerlach
effect in the same way as described in Ref.\,\cite{Mark2007sou}. To do this, we
release the dimers from the trap while the magnetic field gradient is switched
on. After a fixed time of flight, the dimers are dissociated by ramping back
over either the 557.45\,G or the 565.48\,G Feshbach resonance. Subsequently,
the atoms are imaged and the molecular magnetic moment is extracted from the
vertical position of the atom cloud. We observe a smooth change of the magnetic
moment around 557.15\,G over a magnetic field range of about 250\,mG,
indicating that the character of the molecular state also changes smoothly,
from $s$-wave to $i$-wave character, over the width of the crossing. Because of
large calibration uncertainties in these measurements, we cannot provide
absolute values for the molecular magnetic moments. In another experiment, we
start with weakly bound $s$-wave Feshbach molecules at a magnetic field of
560\,G and attempt to jump the avoided crossing diabatically. As in the
experiment previously outlined, we can simultaneously detect and distinguish
$s$-wave and $i$-wave dimers by their magnetic moment. Applying a maximum ramp
speed of 10\,G/ms, the number of transferred dimers is below our detection
limit. This sets a lower limit of 30\,kHz to the strength of the avoided
crossing according to the Landau-Zener formula \cite{Landau1932xxx,
Zener1932xxx}.

\section{Theoretical model}
\label{Theory}

The Hamiltonian for the interaction of two Cs atoms may be written
\begin{equation}
\frac{\hbar^2}{2\mu} \left[-R^{-1} \frac{d^2}{dR^2} R + \frac{\hat
L^2}{R^2} \right] + \hat h_1 + \hat h_2 + \hat V(R),
\label{eq:SE}
\end{equation}
where $\mu$ is the reduced mass and $\hat L^2$ is the operator for the
end-over-end angular momentum of the two atoms about one another. The
monomer Hamiltonians including Zeeman terms are
\begin{equation}
\hat h_j = \zeta \hat \imath_j \cdot \hat s_j + g_e \mu_{\rm B}
B \, \hat s_{zj} + g_n \mu_{\rm B} B \, \hat \imath_{zj},
\label{eq:h-hat}
\end{equation}
where $\zeta$ is the atomic hyperfine constant, $\hat s_1$ and $\hat s_2$
represent the electron spins of the two atoms and $\hat \imath_1$ and $\hat
\imath_2$ represent nuclear spins. $g_e$ and $g_n$ are the electron and nuclear
$g$-factors, $\mu_{\rm B}$ is the Bohr magneton, and $\hat s_z$ and $\hat
\imath_z$ represent the $z$-components of $\hat s$ and $\hat \imath$ along a
space-fixed $Z$ axis whose direction is defined by the external magnetic field
$B$. The interaction between the two atoms $\hat V(R)$ is
\begin{equation}
{\hat V}(R) = \hat V^{\rm c}(R) + \hat V^{\rm d}(R).
\label{eq:V-hat}
\end{equation}
Here $\hat V^{\rm c}(R)=V_0(R)\hat{\cal{P}}^{(0)} + V_1(R)\hat{
\cal{P}}^{(1)}$ is an isotropic potential operator that depends on the
potential energy curves $V_0(R)$ and $V_1(R)$ for the respective
X$^1\Sigma_g^+$ singlet and a$^3\Sigma_u^+$ triplet states of Cs$_2$,
as shown in Figure~\ref{fig-curves}. The singlet and triplet projectors
$\hat{ \cal{P}}^{(0)}$ and $\hat{ \cal{P}}^{(1)}$ project onto
subspaces with total electron spin quantum numbers 0 and 1
respectively. The term $\hat V^{\rm d}(R)$ represents small,
anisotropic spin-dependent couplings that are responsible for the
avoided crossings discussed in the present paper and are discussed
further in Section~\ref{sec:secSO} below.

\subsection{Computational methods for bound states and scattering}

The present work solves the Schr\"odinger equation for both scattering
and Feshbach bound states of Cs$_2$ by coupled-channel methods, using a
basis set for the electron and nuclear spins in a fully uncoupled
representation,
\begin{equation}
|s_1 m_{s1}\rangle|i_1 m_{i1}\rangle
|s_2 m_{s2}\rangle |i_2 m_{i2}\rangle |LM_L\rangle, \label{eqbasdecoup}
\end{equation}
symmetrised to take account of exchange symmetry. The matrix elements
of the different terms in the Hamiltonian in this basis set are given
in the Appendix of Ref.~\cite{Hutson:Cs2-note:2008}. The only
rigorously conserved quantities are the parity, $(-1)^L$, and the
projection of the total angular momentum, $M_{\rm
tot}=m_{s1}+m_{i1}+m_{s2}+m_{i2}+M_L$. The calculations in this paper
used basis sets with all possible values of $m_s$ and $m_i$ for both
atoms that are consistent with the required $M_{\rm tot}$ and parity,
truncated at $L_{\rm max}=4$ (an $sdg$ basis set) unless otherwise
indicated. All calculations in this paper are for $s$-wave incoming
channels, so have even parity.

Both scattering and bound-state calculations use propagation methods
and do not rely on basis sets in the interatomic distance coordinate
$R$.

Scattering calculations are carried out using the MOLSCAT package
\cite{molscat:v14}, as modified to handle collisions in magnetic fields
\cite{Gonzalez-Martinez:2007}. At each magnetic field $B$, the wavefunction
log-derivative matrix at collision energy $E$ is propagated from $R_{\rm
min}=6$ $a_0$ to $R_{\rm mid}=20$ $a_0$ using the propagator of Manolopoulos
\cite{Manolopoulos:1986} with a fixed step size of 0.002 $a_0$, and from
$R_{\rm mid}$ to $R_{\rm max}=4,000$ $a_0$ using the Airy propagator
\cite{Alexander:1987} with a variable step size controlled by the parameter
TOLHI=$10^{-5}$ \cite{Alexander:1984}. Scattering boundary conditions
\cite{Johnson:1973} are applied at $R_{\rm max}$ to obtain the scattering
S-matrix. The energy-dependent $s$-wave scattering length $a(k)$ is then
obtained from the diagonal S-matrix element in the incoming $L=0$ channel using
the identity \cite{Hutson:res:2007}
\begin{equation}
\label{theory:eq5}
a(k) = \frac{1}{ik} \left(\frac{1-S_{00}}{1+S_{00}}\right),
\end{equation}
where $k^2=2\mu E/\hbar^2$.

In the vicinity of a resonance at the lowest atomic threshold, the
scattering length as a function of magnetic field (at small fixed $k$)
follows the functional form
\begin{equation}
\label{theory:eq6}
a(B)=a_{\rm bg}\left[1 - \Delta / (B - B_{\rm res})\right].
\end{equation}
The resonance pole position $B_{\rm res}$ may be associated with the 3-body
loss maximum at a field $B_{\rm max}$. For a narrow resonance (where the
background scattering length $a_{\rm bg}$ does not vary significantly across
the resonance), the width $\Delta$ is conveniently obtained from the difference
between the positions of the pole and zero in $a(B)$. Experimentally, this
corresponds to the difference in field between the loss maximum at $B_{\rm
max}$ and the loss minimum at $B_{\rm min}$. We have extended MOLSCAT to
provide an option to {\em converge} on poles and zeroes of $a(B)$, instead of
extracting them from a fit to a grid of points.

Weakly bound levels for Feshbach molecules are obtained using a variant of the
propagation method described in Ref.\ \cite{Hutson:Cs2-note:2008}. The
log-derivative matrix is propagated outwards from $R_{\rm min}$ to $R_{\rm
mid}$ with a fixed step size of 0.002 $a_0$ and inwards from $R_{\rm max}$ to
$R_{\rm mid}$ with a variable step size, using the same propagators as for
scattering calculations. $R_{\rm mid}=25$ $a_0$ and $R_{\rm max}=4,000$ $a_0$
were used for most bound states, although $R_{\rm mid}=35$ $a_0$ and $R_{\rm
max}=8,000$ $a_0$ were needed for states within about 50 kHz of dissociation.
In Ref.~\cite{Hutson:Cs2-note:2008}, bound-state energies at a fixed value of
the magnetic field $B$ were located using the BOUND package
\cite{Hutson:bound:1993}, which converges on energies where the smallest
eigenvalue of the log-derivative matching determinant is zero
\cite{Hutson:CPC:1994}. However, for the purposes of the present work we used a
new package, FIELD \cite{Hutson:field:2011}, which instead works at fixed
binding energy and converges in a similar manner on the magnetic fields at
which bound states exist. BOUND and FIELD both implement a node-count algorithm
\cite{Hutson:CPC:1994} which makes it straightforward to ensure that {\em all}
bound states that exist in a particular range of energy or field are located.

As described above, zero-energy Feshbach resonances can in principle be located
as the fields at which the scattering length $a(B)$ passes through a pole.
However, with this method it is necessary first to search for poles, and it is
quite easy to miss narrow resonances. However, since resonances occur at fields
where there is a bound state at zero energy, the FIELD package provides a much
cleaner approach: simply running FIELD at zero energy provides a complete list
of all fields at which zero-energy Feshbach resonances exist \cite{SuppMat}. In the present
work we located resonances using FIELD and then obtained their widths by
converging on the nearby zero in $a(B)$ using MOLSCAT.

\subsection{Representation of the potential curves}

At long range, the potentials are
\begin{eqnarray}
V_S^{\rm LR}(R) = &-& C_6y_6(R)/R^6 - C_8y_8(R)/R^8
\nonumber\\
&-& C_{10}y_{10}(R)/R^{10} \pm V_{\rm ex}(R),
\label{eq:vlr}
\end{eqnarray}
where $S=0$ and 1 for singlet and triplet, respectively. The dispersion
coefficients $C_n$ are common to both potentials and the functions
$y_n(R)$ account for retardation corrections \cite{Marinescu1994}. The
exchange contribution is \cite{Smirnov:1965}
\begin{equation}
V_{\rm ex}(R) = A_{\rm ex} (R/a_0)^\gamma \exp(-\beta_{\rm ex} R/a_0),
\end{equation}
and makes an attractive contribution for the singlet and a repulsive
contribution for the triplet. The value of $\beta_{\rm ex}$ is usually
obtained from the ionization energies of the atoms \cite{Smirnov:1965},
which for Cs gives $\beta_{\rm ex}=1.069946$, and $\gamma$ is related
to $\beta$ by $\gamma=7/\beta-1$. In the present work we found that, to
reproduce the experimental results, it was necessary to reduce
$\beta_{\rm ex}$ slightly from its original value. We therefore
introduce an additional factor $\rho_{\rm ex}$ so that $\beta_{\rm
ex}=1.069946\rho_{\rm ex}$, with $\gamma$ adjusted accordingly.

The detailed shapes of the short-range singlet and triplet potentials
are relatively unimportant for the ultracold scattering properties and
near-threshold binding energies considered here, although it is crucial
to be able to vary the {\em volume} of the potential wells to allow
adjustment of the singlet and triplet scattering lengths. In the
present work we retained the functional form used by Leo {\em et al.}\
\cite{Leo2000cpo} and Chin {\em et al.}\ \cite{Chin2004pfs}. Each
short-range potential is represented by a set of 14 ab initio points
between $R=7$ and 20 $a_0$ \cite{Krauss1990}. The two sets of potential
points are first multiplied by $R^6$ and the resulting (smoother)
functions are interpolated using Akima splines \cite{Akima1991} to
obtain their values at $R_{\rm LR}=17.6$ $a_0$. The value of $A_{\rm
ex}$ is chosen to match $V_{\rm ex}$ to $(V_1-V_0)/2$ at $R_{\rm LR}$,
and both sets of points are shifted to match $(V_1+V_0)/2$ at $R_{\rm
LR}$. Finally, the analytic $V_S^{\rm LR}(R)$ is used to generate new
grid points between 17.6 $a_0$ and 20 $a_0$ and the resulting sets of
points are reinterpolated as above between $R=7$ and 20 $a_0$. The
analytic long-range form (\ref{eq:vlr}) is used outside 20 $a_0$.

The flexibility needed to adjust the singlet and triplet scattering
lengths is provided by simply adding a quadratic shift to each of the
singlet and triplet potentials inside its minimum,
\begin{equation}
V_S^{\rm shift}(R) = S_S (R-R_{eS})^2 \hbox{\qquad for\qquad} R<R_{eS},
\end{equation}
with $R_{e0}=8.75$ $a_0$ and $R_{e1}=11.8$ $a_0$.

\subsection{Magnetic dipole interaction and second-order spin-orbit
coupling} \label{sec:secSO}

At long range, the coupling $\hat V^{\rm d}(R)$ of Eq.\
(\ref{eq:V-hat}) has a simple magnetic dipole-dipole form that varies
as $1/R^3$~\cite{Stoof:1988, Moerdijk:1995}. However, for atoms as
heavy as Cs, second-order spin-orbit coupling provides an additional
contribution that has the same tensor form as the dipole-dipole term
and dominates at short range \cite{Mies:1996}. In the present work,
$\hat V^{\rm d}(R)$ is represented as
\begin{equation}
\label{eq:Vd} \hat V^{\rm d}(R) = \lambda(R) \left ( \hat s_1\cdot
\hat s_2 -3 (\hat s_1 \cdot \vec e_R)(\hat s_2 \cdot \vec e_R)
\right ) \,,
\end{equation}
where $\vec e_R$ is a unit vector along the internuclear axis and
$\lambda$ is an $R$-dependent coupling constant. The second-order term
has been calculated by Kotochigova {\em et al.} \cite{Kotochigova:2001}
and fitted to a biexponential form, so that the overall form of
$\lambda(R)$ is
\begin{eqnarray}
\label{eq:lambda}
\lambda(R) = E_{\rm h} \alpha^2 \bigg[
&A_{\rm 2SO}^{\rm short}& \exp\left(-2\beta_{\rm 2SO}R\right)
\nonumber\\
+ &A_{\rm 2SO}^{\rm long}&
\exp\left(-\beta_{\rm 2SO}R\right)
+  \frac{1}{(R/a_0)^3}\bigg],
\end{eqnarray}
where $\alpha\approx 1/137$ is the fine-structure constant and the
parameters obtained from fitting to the electronic structure
calculations \cite{Kotochigova:2001} are $A_{\rm 2SO}^{\rm short}/hc =
34.4$ cm$^{-1}$, $A_{\rm 2SO}^{\rm long}/hc = 0.25$ cm$^{-1}$ and
$\beta_{\rm 2SO} = -0.35$ $a_0^{-1}$. However, in fitting to the
experimental results, this coupling function was found to be slightly
too strong. We therefore retained the functional form (\ref{eq:lambda})
but introduced an additional scaling factor $S_{\rm 2SO}$ that
multiplies both exponential terms and is allowed to vary in the
least-squares fit to the experimental results.

\section{Least-squares fitting of potential parameters}
\label{Fit}

In the present work, our primary objective is to obtain potential
parameters that give a reliable representation of $a(B)$ in the regions
where Efimov resonances occur, namely near 8 G, 554 G and 853 G.  Earlier
potentials \cite{Leo2000cpo, Chin2004pfs} focused on representing
the positions of Feshbach resonances in the low-field region below
about 60~G.

A key advantage of the propagator approach to locating bound states and
resonances, implemented in the BOUND and FIELD programs, is that it is
fast enough to be incorporated in a least-squares fitting program. We
have therefore carried out direct least-squares refinement of the
potential parameters. We experimented with fitting various combinations
of parameters, and concluded that adequate flexibility is available in
the 6-parameter space $S_0$, $S_1$, $C_6$, $C_8$, $S_{\rm 2SO}$,
$\rho_{\rm ex}$.

\begin{table*}[htpb]
\caption{Quality of fit between calculations using the M2012 model and the
experimental results used in the fit.} \label{quality-of-fit}
\begin{tabular}{l r r r c c c}
\hline\hline
& $B_{\rm obs}$ (G) & $B_{\rm calc}$ (G) & $B_{\rm obs}-B_{\rm calc}$ (G) & Unc. (G) & method & reference\\
\hline
$-7(44)6s(6)$ at 7.8 MHz                    &   17.53 &   17.51 &    0.02~~~~~~~~&   0.02 & microwave spectr. & \cite{Mark2007sou}\\
$-7(44)6s(6)$ at 1.2 MHz                    &   21.60 &   21.59 &    0.01~~~~~~~~&   0.02 & microwave spectr. & \cite{Mark2007sou}\\
$-7(44)6s(6)$ at 104 kHz                    &   32.05 &   31.70 &    0.35~~~~~~~~&   0.03 & magnetic field mod. & \cite{Lange2009doa}\\
Zero crossing near 17 G                  &   17.12 &   17.14 &  $-0.02$~~~~~~~~&   0.01 & Bloch oscillations & \cite{Gustavsson2008coi}\\
$-2(33)4d(4)$ at 174 kHz                 &   48.01 &   48.01 &    0.00~~~~~~~~&   0.06 & magnetic field mod.  & \cite{Lange2009doa}\\
$-2(33)4d(4)$ crossing strength 78-24 kHz&   1.19  &    1.21 &  $-0.02$~~~~~~~~&   0.02 & magnetic field mod.  & \cite{Hutson:Cs2-note:2008}\\
Loss minimum ($d$) near 48 G             &   47.94 &   47.98 &  $-0.04$~~~~~~~~&   0.04 & inferred from magn. field mod. & \cite{Lange2009doa} \\
Loss maximum ($d$) near 48 G             &   47.78 &   47.79 &  $-0.01$~~~~~~~~&   0.06 & inferred from magn. field mod. & \cite{Lange2009doa} \\
$\Delta\ (d)$ near 48 G                  &    0.16 &    0.18 &  $-0.02$~~~~~~~~&   0.06 & inferred from magn. field mod. & \cite{Lange2009doa} \\
$2g(2)$ at 17 kHz                        &   53.42 &   53.76 &  $-0.34$~~~~~~~~&   0.08 & magnetic field mod.  & \cite{Lange2009doa}\\
$-2(33)6g(6)$ at 18.6 G (MHz, not G)     &  $-5.03$& $-4.99$ &  $-0.04$~~~~~~~~&   0.01 & microwave spectr. & \cite{Mark2007sou}\\
\hline
Loss maximum $-6(34)7d(x)$               &  492.45 &  492.68 &  $-0.23$~~~~~~~~&   0.06 & trap loss spectr. & {\it this work}\\
Loss maximum $-6(34)7d(x)$               &  501.24 &  501.44 &  $-0.20$~~~~~~~~&   0.06 & trap loss spectr. & {\it this work}\\
Loss maximum $-6(34)7d(x)$               &  505.07 &  505.37 &  $-0.30$~~~~~~~~&   0.06 & trap loss spectr. & {\it this work}\\
$-6(34)7s(6)$ at 1.0 MHz                 &  556.47 &  556.48 &  $-0.01$~~~~~~~~&   0.02 & magnetic field mod. & {\it this work} \\
$-6(34)7s(6)$ at 700 kHz                 &  556.72 &  556.76 &  $-0.04$~~~~~~~~&   0.02 & magnetic field mod. & {\it this work}\\
$-6(34)7s(6)$ at 170 kHz                 &  557.88 &  557.80 &    0.08~~~~~~~~&   0.03 & magnetic field mod. & {\it this work}\\
$-6(34)7s(6)$ at 100 kHz                 &  558.44 &  558.36 &    0.08~~~~~~~~&   0.03 & magnetic field mod. & {\it this work}\\
Zero crossing near 556 G                 &  556.26 &  556.19 &    0.07~~~~~~~~&   0.03 & collapse of BEC & \cite{Zenesini2012cob} \\
$-3(33)6g(3)$ at 368 kHz                 &  553.44 &  553.44 &  $-0.00$~~~~~~~~&   0.01 & magnetic field mod. & {\it this work}\\
Loss minimum ($g$) near 554 G            &  553.73 &  553.75 &  $-0.02$~~~~~~~~&   0.01 & trap loss spectr. & {\it this work}\\
Loss maximum ($g$) near 554 G            &  554.06 &  554.07 &  $-0.01$~~~~~~~~&   0.02 & trap loss spectr. & {\it this work}\\
$\Delta\ (g)$ near 554 G                 &    0.33 &    0.32 &    0.01~~~~~~~~&   0.01 & trap loss spectr. & {\it this work}\\
$-6(34)6s(6)$ at 1.7 MHz                 &  890.52 &  890.61 &  $-0.09$~~~~~~~~&   0.02 & magnetic field mod. & {\it this work}\\
$-6(34)6s(6)$ at 356 kHz                 &  899.93 &  900.19 &  $-0.26$~~~~~~~~&   0.03 & magnetic field mod. & {\it this work}\\
$-6(34)6s(6)$ at 110 kHz                 &  915.66 &  915.54 &    0.12~~~~~~~~&   0.03 & magnetic field mod. & {\it this work}\\
Zero crossing near 881 G                 &  880.90 &  880.66 &    0.24~~~~~~~~&   0.03 & collapse of BEC & \cite{Zenesini2012cob}\\
$-6(34)6d(6)$ at 342 kHz                 &  819.17 &  819.20 &  $-0.03$~~~~~~~~&   0.03 & magnetic field mod. & {\it this work}\\
Loss minimum (d) near 820 G              &  819.41 &  819.37 &    0.04~~~~~~~~&   0.03 & trap loss spectr. & {\it this work}\\
Loss maximum (d) near 820 G              &  820.37 &  820.33 &    0.04~~~~~~~~&   0.02 & trap loss spectr. & {\it this work}\\
$\Delta\ (d)$ near 820 G                 &    0.96 &    0.97 &  $-0.01$~~~~~~~~&   0.05 & trap loss spectr. & {\it this work}\\
\hline
\end{tabular}
\end{table*}

The set of experimental results used for fitting is listed in Table
\ref{quality-of-fit}. It consists of all the observed resonance
positions (loss maxima) in the regions relevant to Efimov physics,
together representative bound-state positions. It also includes zero
crossings of the scattering lengths measured from loss minima, Bloch
oscillations \cite{Gustavsson2008coi} and BEC collapse measurements
\cite{Zenesini2012cob}. All the bound-state positions are expressed as
the fields at which bound states exist at specific binding energies,
except for the $-3(33)6g(6)$ state, which is almost parallel to the
lowest threshold and is included in the fit as a binding energy at
$B=18.6$ G. The strength of the avoided crossing between the
$-2(33)4d(4)$ and $-7(44)6s(6)$ states near 48 G is included as an
explicit difference between fields where states exist at energies of 78
kHz and 24 kHz. The quantity optimized in the least-squares fits was
the sum of squares of residuals ((obs$-$calc)/uncertainty), with the
uncertainties listed in Table \ref{quality-of-fit} \cite{Note2}.

\begin{table}[htpb]
\caption{Parameters of the fitted potential.} \label{fitparms}
\begin{tabular}{l|rrr}
\hline\hline
& fitted value & confidence   & sensitivity \\
&              & limit (95\%) &             \\
\hline
$C_6$ ($E_{\rm h}a_0^6$)    & 6890.4768                & 0.081                & 0.0003 \\
$C_8$ ($E_{\rm h}a_0^8$)    & 1009289.6                & 2900                 & 0.1    \\
$S_0$ ($E_{\rm h}a_0^{-2}$) & $3.172749\times 10^{-4}$ & $2.5\times 10^{-6}$  & $2.7\times 10^{-10}$ \\
$S_1$ ($E_{\rm h}a_0^{-2}$) & $1.343217\times 10^{-4}$ & $2.5\times 10^{-6}$  & $1.3\times 10^{-10}$ \\
$S_{\rm 2SO}$               & 1.36432                  & 0.022                & 0.00065 \\
$\rho_{\rm ex}$             & 0.978845                 & 0.0026               & $5.5\times 10^{-7}$ \\
\hline\hline
\end{tabular}
\end{table}
Although the experimental data do allow all 6 of the potential parameters
described above to be determined, the fit is very highly correlated. Under
these circumstances, a fully automated approach to fitting is unreliable:
individual least-squares steps often reach points in parameter space where the
levels have moved too far to be identified reliably, particularly in the early
stages of fitting. We therefore carried out the fitting using the I-NoLLS
package \cite{I-NoLLS} (Interactive Non-Linear Least-Squares), which gives the
user interactive control over step lengths and assignments as the fit proceeds.
This allowed us to converge on a minimum in the sum of weighted squares, with
the parameters given in Table \ref{fitparms}.

\begin{table}[htpb]
\caption{Comparison between key quantities calculated from different
potentials.} \label{derived}
\begin{tabular}{l|rr}
\hline\hline
derived    & M2012  & M2004 \\
parameters &       &      \\
\hline
$a_S$ (bohr) & 286.5(1) & 280.37(6)  \\
$a_T$ (bohr) & 2858(19)  & 2440(24)  \\
$s$ pole near $-$10 (G) & $-$12.38(8) &    \\
$s$ pole near 550 (G) & 548.78(9) &    \\
$s$ pole near 800 (G) & 786.8(6) &   \\
\hline\hline
\end{tabular}
\end{table}

The parameter uncertainties are given in Table \ref{fitparms} as 95\%
confidence limits \cite{LeRoy:1998}. However, it should be emphasized that
these are {\em statistical} uncertainties within the particular parameter set.
They do not include any errors due to the choice of the potential functions.
Such model errors are far harder to estimate, except by performing a large
number of fits with different potential models, which is not possible in the
present case.

In a correlated fit, the statistical uncertainty in a fitted parameter
depends on the degree of correlation. However, to reproduce the results
from a set of parameters, it is often necessary to specify many more
digits than implied by the uncertainty. A guide to the number of digits
required is given by the {\em parameter sensitivity}
\cite{LeRoy:1998}, which essentially measures how fast the observables
change when one parameter is varied with all others held fixed. This
quantity is included in Table \ref{fitparms}.

The singlet and triplet scattering lengths and the pole positions of
the $s$-wave resonances are not directly observed quantities.
Nevertheless, their values may be extracted from the final potential.
In addition, the statistical uncertainties in derived parameters such
as these may be obtained as described in Ref.\ \onlinecite{LeRoy:1998}.
The values and 95\% uncertainties obtained in this way are give in
Table \ref{derived}.

\subsection{Region between 800 G and 920 G}

\begin{figure}[htbp]
\includegraphics[width=\columnwidth]{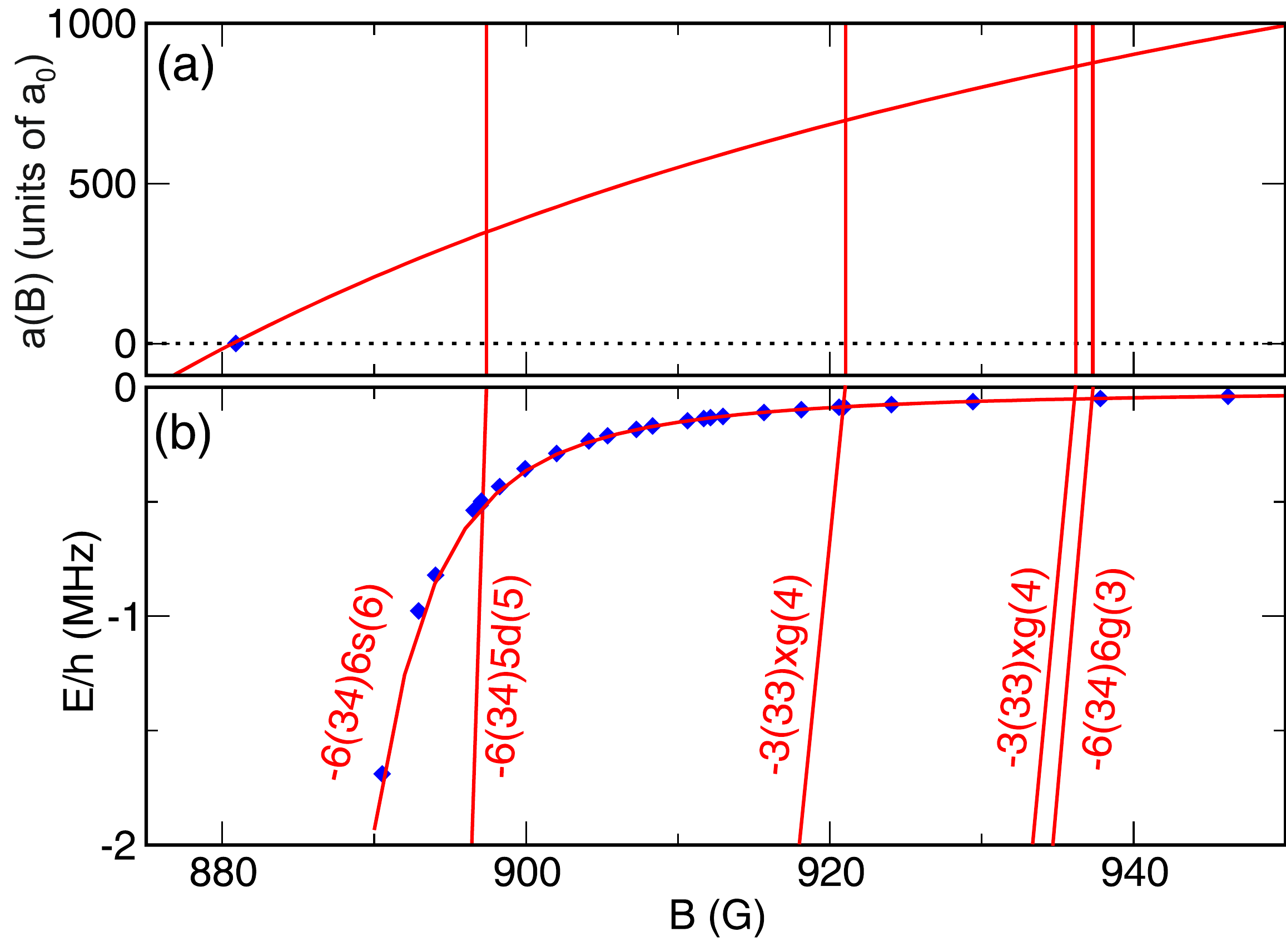}
\caption{(Color online) (a) The scattering length in the region between
875 G and 950 G. The solid lines show results calculated for the M2012
potential with an $sdg$ basis set. The diamond near 880 G
indicates the position of the zero crossing. (b) The calculated
bound-state energies with the same basis set. The diamonds show
the measured energies in this region.} \label{s900}
\end{figure}

\begin{figure}[htbp]
\includegraphics[width=\columnwidth]{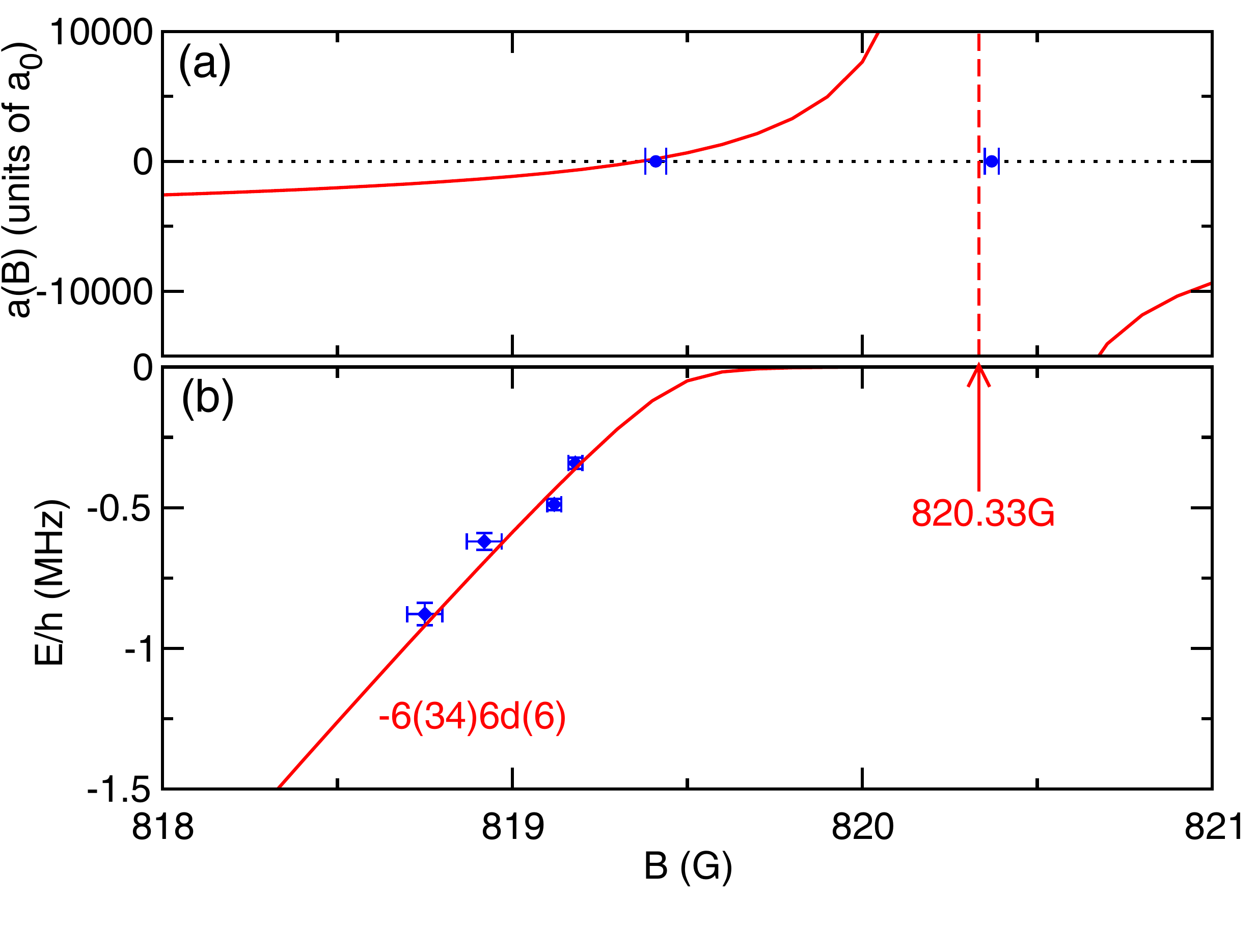}

\caption{(Color online) The scattering length and bound-state energy in
the pole region of the $-6(34)6d(6)$ resonance. The solid lines show
results from model M2012, calculated with an $sdg$ basis set. The points in panel (a) show the measured pole position and loss minimum. The points in panel (b) show the measured binding energies. The dashed
line in (a) and the arrow in (b) shows the calculated pole position. Error bars refer to $1\sigma$ uncertainties.}
 \label{d820}
\end{figure}

In the region between 800 and 920~G, the near-threshold molecular structure and the corresponding scattering properties are relatively straightforward. As seen
in Fig.\ \ref{figsd6}, the ramping $-6(34)6s(6)$ state has a strong
avoided crossing with $-2(33)6s(6)$ near 850 G at a binding energy of
about 110 MHz. This crossing is still incomplete when the mixed state
crosses the least-bound state $-1(33)6s(6)$ around 100 kHz. It is the
state resulting from this second crossing that is observed at binding
energies of 0.1 to 1.7 MHz between 890 and 920 G. These points are very
well reproduced by the fit, as seen in Figure \ref{s900}. The
$-3(34)6d(6)$ state that crosses the axis at 820.37 G is also well
reproduced by the fit, as seen in Figure \ref{d820}.

The position of the $s$-wave resonance pole in this region has not been
observed because of large 3-body losses, but it is predicted by the fitted
potential at 786.8(6) G. It is interesting to note that, because of the shift
produced by the crossing with the least-bound state, the $s$-wave resonance
occurs at a {\em lower} field than the corresponding $d$-wave resonance in this
case, even though (as always) the unperturbed ramping $s$-wave state is lower
in energy.

\subsection{Region around 550 G}

The region around 550 G is considerably more complicated. In this case
the crossing between the ramping $-6(34)7s(6)$ state and $-2(33)6s(6)$
near 510 G is much narrower, so is almost complete by the time
$-6(34)7s(6)$ crosses the least-bound state $-1(33)6s(6)$. This
produces a zero crossing near 556 G. However, as described in the
experimental section, there is also a ramping $-3(33)3g(3)$ state that
crosses threshold just below this, producing an additional pole and
zero crossing near 554 G. Because of the large background scattering
length arising from its proximity to the $s$-wave resonance, the
$g$-wave resonance is far wider than is usual.

The bound-state measurements near 557 G are further complicated by an
additional state that crosses and appears to mix with the $s$-wave state in the
top 1~MHz. This cannot be assigned as $s$-, $d$- or $g$-wave; it must be due to
an $i$-wave state, and indeed the M2004 model predicts four $i$-wave levels to
be in the range between around 530~G and 590~G, with two between 555~G and
565~G. However, these states are associated with hyperfine-excited thresholds,
and their exact positions are very sensitive to details of the potential that
do not significantly affect the other experimentally measured quantities
considered here. We have been unable to decide unambiguously which one is
responsible for the observed crossing. In addition, the $i$-wave states have
very little influence on the $s$-wave scattering length that is the main object
of interest in this region. We therefore decided to fit using a basis set with
$L_{\rm max}=4$, which excludes the $i$-wave states entirely, and also to
exclude from the fit any binding energies affected by the crossing between the
$s$- and $i$-wave states.

\begin{figure}[tbp]
\includegraphics[width=\columnwidth]{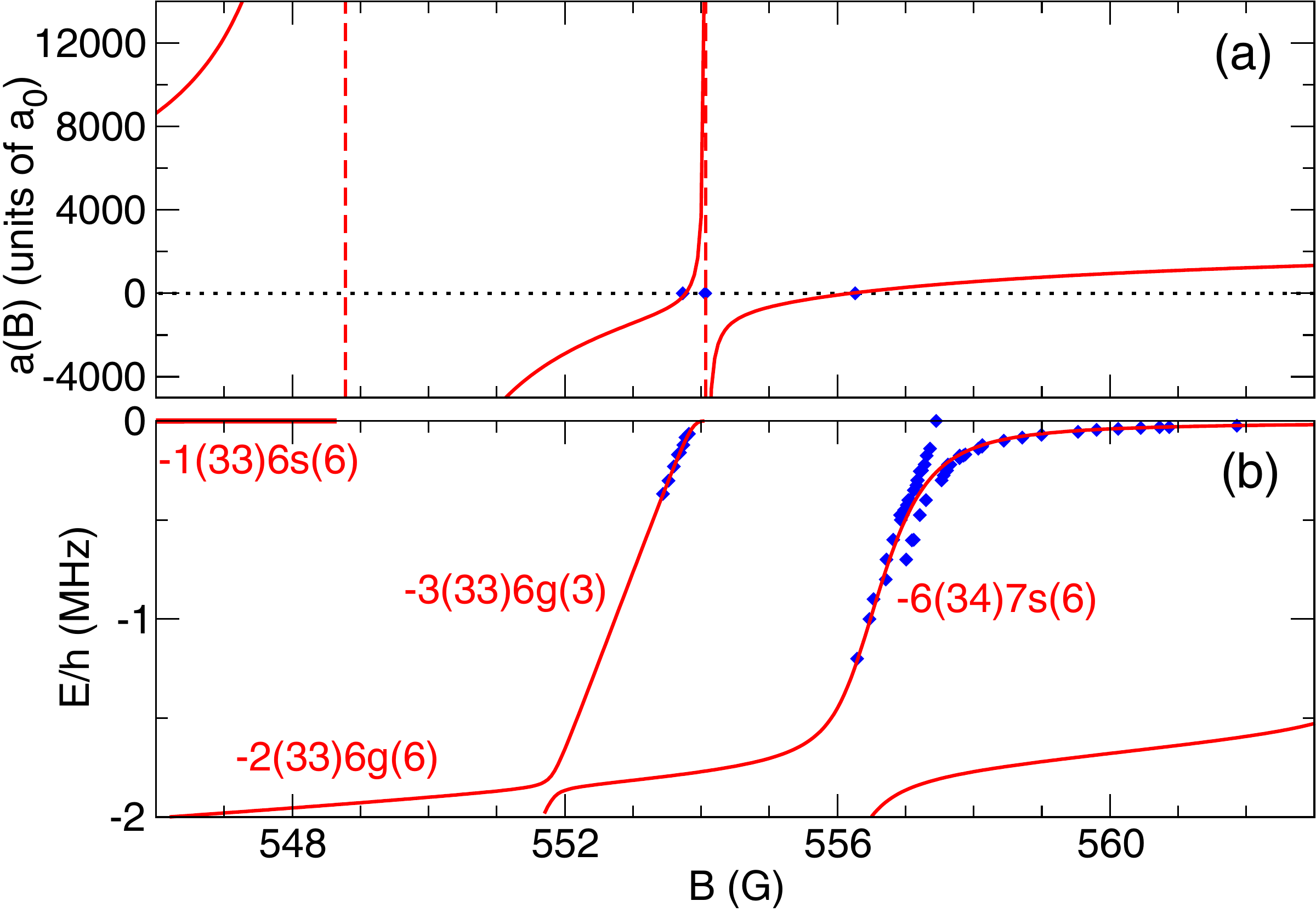}
\caption{(Color online) (a) The scattering length between 546 G and 564
G. The solid lines show results calculated for the M2012 potential with
an $sdg$ basis set. The diamonds show the measured pole position
and loss minimum near 554 G and the zero crossing near 556 G. The
dashed line shows the calculated pole position. (b) The calculated
binding energies with the same basis set as for Panel (a). The diamonds show the measured binding energies. } \label{s556}
\end{figure}

\begin{figure}[htbp]
\includegraphics[width=\columnwidth]{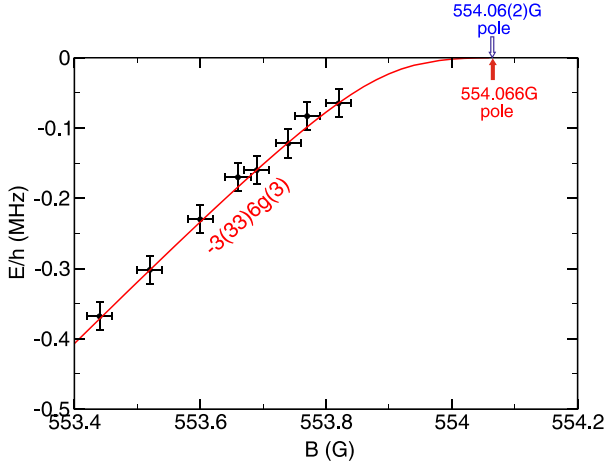}
\caption{(Color online) Expanded view of the scattering length in the
pole region of the $-3(33)6g(3)$  resonance. The solid line shows
results calculated for the M2012 potential with an $sdg$ basis set. The
points show the measured bound-state energies. The calculated and
measured pole positions are indicated by solid and open arrows, respectively. Error bars refer to $1\sigma$ uncertainties.} \label{g554}
\end{figure}

The general fit to the ramping $s$-wave bound state is shown in Fig.\
\ref{s556} and an expanded view of the fir to the $g$-wave state is shown in
Fig.\ \ref{g554}. It may be seen that the calculated $s$-wave state passes well
through the experimental points either side of the crossing with the $i$-wave
state, while the calculated $g$-wave state reproduces the bound-state energies
as well as the zero crossing and pole in the scattering length.

The crossing between the $i$-wave state and the $-6(34)7s(6)$ state
near 557 G is surprisingly strongly avoided. As described above, we
were unable to ramp the magnetic field fast enough to transfer a
detectable number of molecules between the two states, which sets a
lower limit of 30 kHz to the strength of the avoided crossing. In a
zeroth-order picture, the two states are separated by $\Delta L=6$ and
the only coupling off-diagonal in $L$ is $\hat V^{\rm d}(R)$, which can
couple only $\Delta L=2$ and is quite small. However, quantitative
modeling of this effect requires a theoretical model that places the $i$-wave
state in the correct place. This is likely to require simultaneous
modeling of the present results and the spectroscopy of more deeply
bound levels, and this remains a subject for future work.

\subsection{Low-field region}

The low-field region, below 60 G, is also quite complicated. The ramping
$s$-wave state responsible for the Efimov resonances in this region is
$-7(44)6s(6)$. However, there are also families of ramping $-2(33)6g$ states
that cross threshold between 4 and 8 G and $-2(33)4g$ states that cross between
11 and 21 G. In addition, there is a $-2(33)4d(4)$ state that crosses near 48 G
and a $2g(2)$ state that crosses near 54 G. The latter state does not carry a
clear $n(f_1f_2)$ signature, but it clearly has $2g(2)$ character.

The binding energies of many of these states have been measured by
magnetic moment spectroscopy at binding energies up to about 10 MHz
\cite{Mark2007sou}, though these measurements have significant
uncertainties associated with the integration over field. However,
there are much more precise results for the $-7(44)6s(6)$ state as it
crosses with the least-bound state, obtained from microwave
spectroscopy \cite{Mark2007sou} and magnetic field modulation
spectroscopy \cite{Lange2009doa}. Particularly important are
measurements of the crossing between this mixed state and the
$-2(33)4d(4)$ state near 48~G, since the strength of this crossing
provides the most direct experimental information available on the
strength of the second-order spin-orbit coupling and thus on the
potential parameter $S_{\rm 2SO}$.

\begin{figure}[htbp]
\includegraphics[width=\columnwidth]{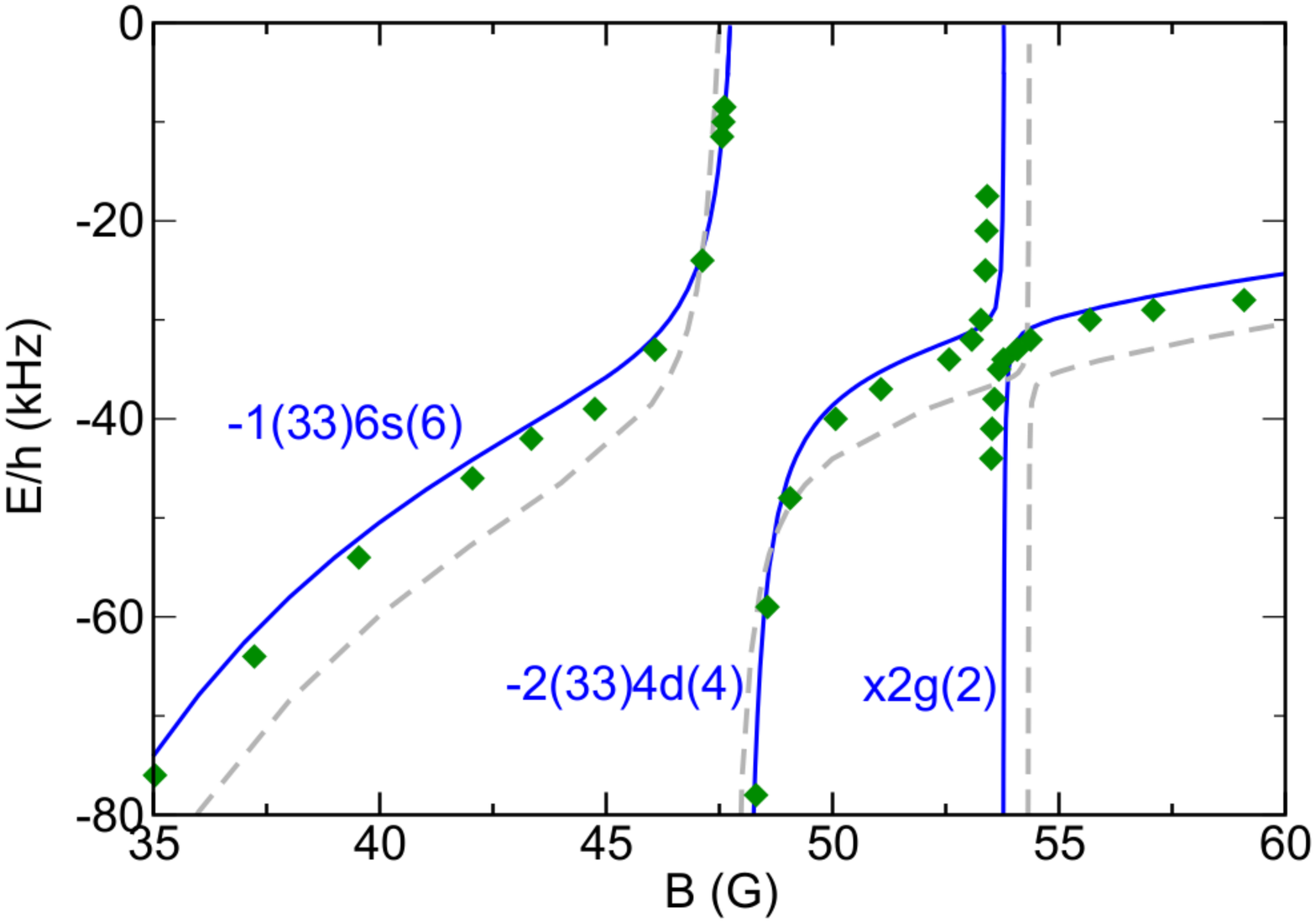}
\caption{(Color online) The bound-state energies between 35 G and 60 G.
The solid lines show results calculated for the M2012 potential with an
$sdg$ basis set, showing one side of the avoided crossings of the
$-7(44)6s(6)$ bound state with the $-1(33)6s(6)$ least-bound state in
the threshold entrance channel. The points show the previously measured
bound-state energies~\cite{Mark2007sou,Lange2009doa}.  The agreement is
much better with the new M2012 model than with the old M2004 model,
shown as light dashed lines.} \label{d46}
\end{figure}

\begin{figure}[htbp]
\includegraphics[width=\columnwidth]{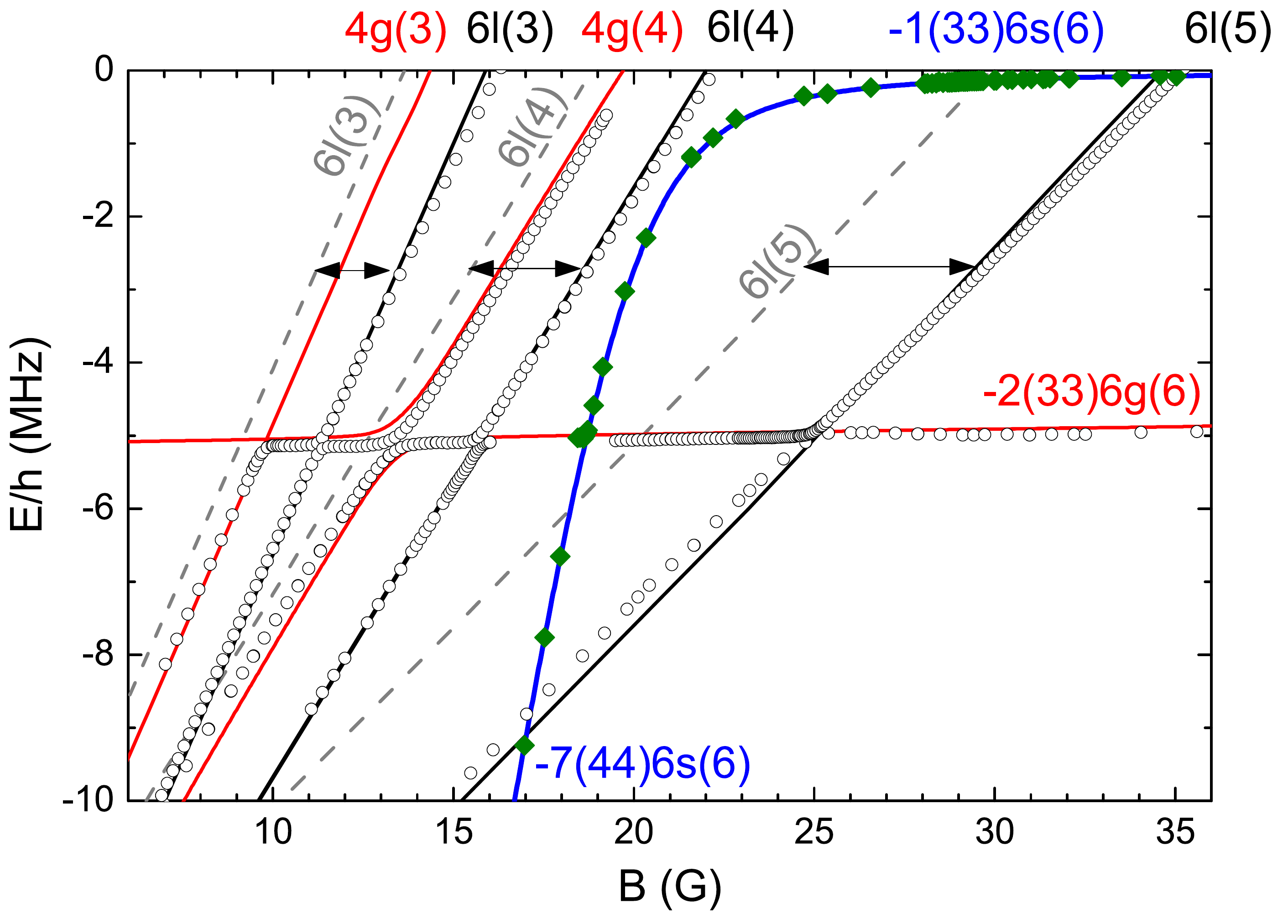}
\caption{(Color online) Comparison between experimental results for $s$-wave
levels (diamonds ~\cite{Mark2007sou,Lange2009doa}) and $g$- and $l$-wave
levels (open circles~\cite{Mark2007sou, Knoop2008mfm}) with the results of the
M2012 model (solid lines). Only levels for which experimental results exist are
shown. The $s$-wave and $g$-wave levels are calculated with the full $sdg$
basis set. The $l$-wave levels are calculated with an $l$ basis only. The light
dashed lines show the $l$-wave levels calculated with the M2004 model. The
arrows show the separation between equivalent levels in the M2004 and M2012
models. The additional quantum numbers $n(f_1f_2)$ are $-2(33)$ for the two
$4g$ levels and $-3(33)$ for the three $6l$ levels.} \label{L20}
\end{figure}

Figures~\ref{d46} and~\ref{L20} show the overall fit to the bound
states below 60~G for both the M2004 and M2012 potentials. The ramping
$s$-wave state between 17 G and 60 G is of particular interest. This
state switches from $-7(44)6s(6)$ to $-1(33)6s(6)$ character as $B$
increases. All the experimental energies are well reproduced, including
those near the two avoided crossings with the $4d(4)$ and $2g(2)$
states shown in Fig.~\ref{d46}. Even in this region, the M2012 model
agrees with the experimental results significantly better than the
M2004 model.

\subsection{Independent tests of the M2012 model}

The older potential models \cite{Leo2000cpo, Chin2004pfs} satisfactorily
reproduced the bound states for $s$-, $d$- and $g$-wave states at fields below
60 G. However, the measurements also revealed the existence of $l$-wave ($L =
8$) states \cite{Mark2007sou, Knoop2008mfm}, which do not lead to observable
Feshbach resonances because of their very weak coupling to the $s$-wave
threshold channel. The M2004 model failed for these $l$-wave states, with
errors of up to 5~G in the positions at a given bound-state energy. The
$l$-wave states were not included explicitly in our fits, but the comparison
between the calculated levels and experiment is shown in Fig.\ \ref{L20} for
both the current model M2012 and the older M2004 one. It is clear that M2012
gives a far more satisfactory reproduction of the experimental $l$-wave levels.

\begin{figure}[tbp]
\includegraphics[width=\columnwidth]{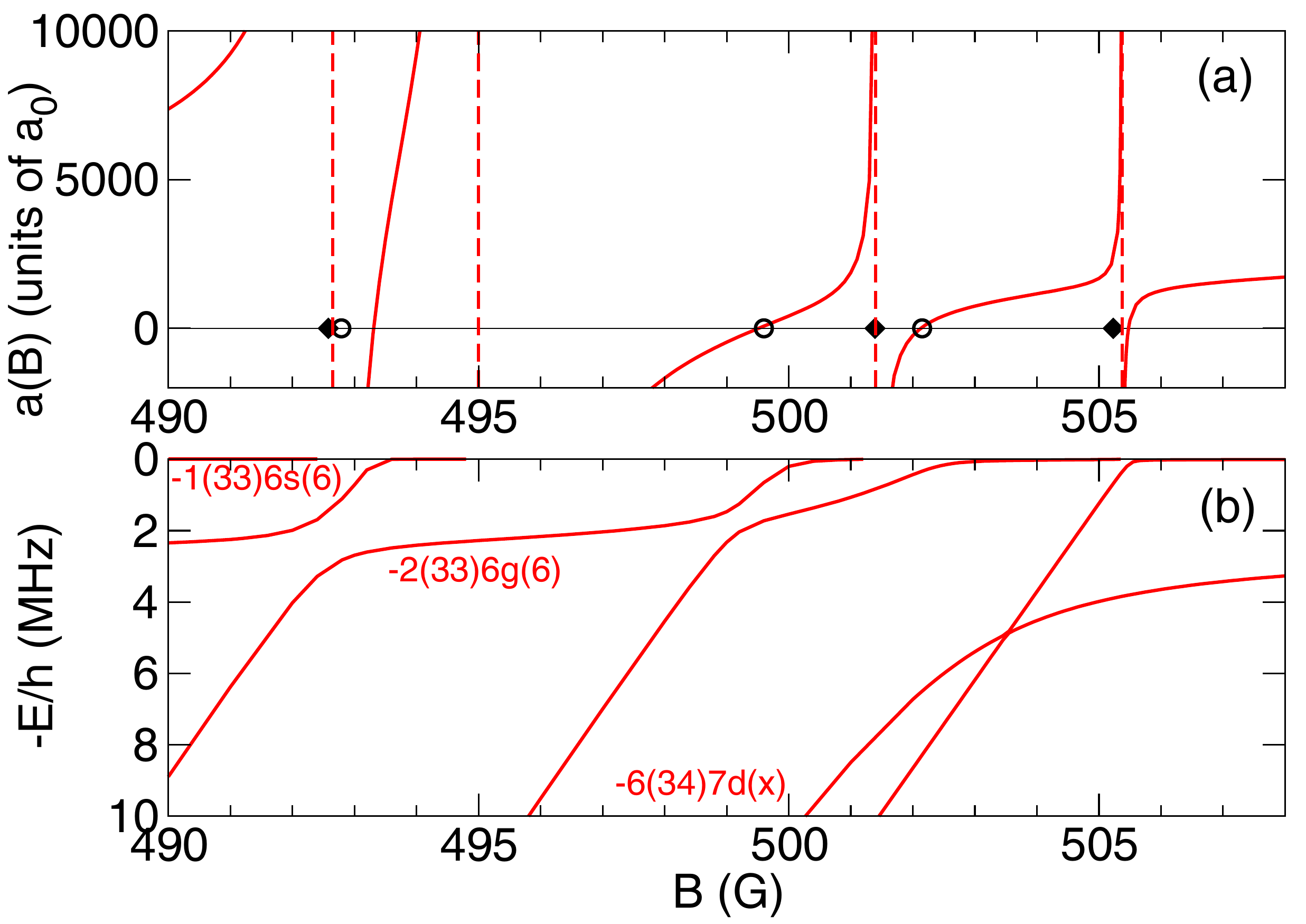}
\caption{(Color online) The scattering length (a) and bound-state
energies (b) between 490 G and 508 G, calculated for the M2012
potential with an $sdg$ basis set. The solid diamonds in (a) show the
positions of experimental loss maxima and the open circles show the
positions of loss minima. The four ramping $-6(34)7d(x)$ levels,
labeled collectively by $x$ since their $M_F$ components are mixed,
undergo avoided crossings with the $-2(33)6g(6)$ level that passes
through this region with $-E/h\approx 2$ MHz. The avoided crossings
with the least-bound level $-1(33)6s(6)$ near $-E/h \approx 0.01$ MHz
are too close to the $E=0$ line to be seen on the figure. }
\label{g500}
\end{figure}

There is a particularly interesting region near 500 G, where a group of
four strongly coupled $-6(34)7d(M_F)$ levels cross threshold. The
underlying levels have $M_F$ values of 4, 5, 6, and 7, but are strongly
mixed with one another so that $M_F$ is not a good quantum number for
the actual eigenstates. The bound states have not yet been measured in
this region, but the calculated levels are shown in Fig.\ \ref{g500},
together with the calculated scattering length and the positions of the
measured loss maxima and minima. The loss maxima are well reproduced.
The two loss minima at 499.6 G and 502.15 G are also well reproduced by
the model, but the loss minimum at 492.8 G is not near a zero-crossing
of the calculated scattering length.  However, there are two strong
overlapping and interfering resonances with poles in $a(B)$ at 492.7 G
and 495.0 G, and it is not clear how to interpret the three-body loss
in such a region. This complex region from 490 G to 510 G needs further
investigation, especially since it may display rich Efimov physics
\cite{Ferlaino2011eri}.

\subsection{Mapping between scattering length and magnetic field}

\begin{figure}[htbp]
\includegraphics[width=\columnwidth]{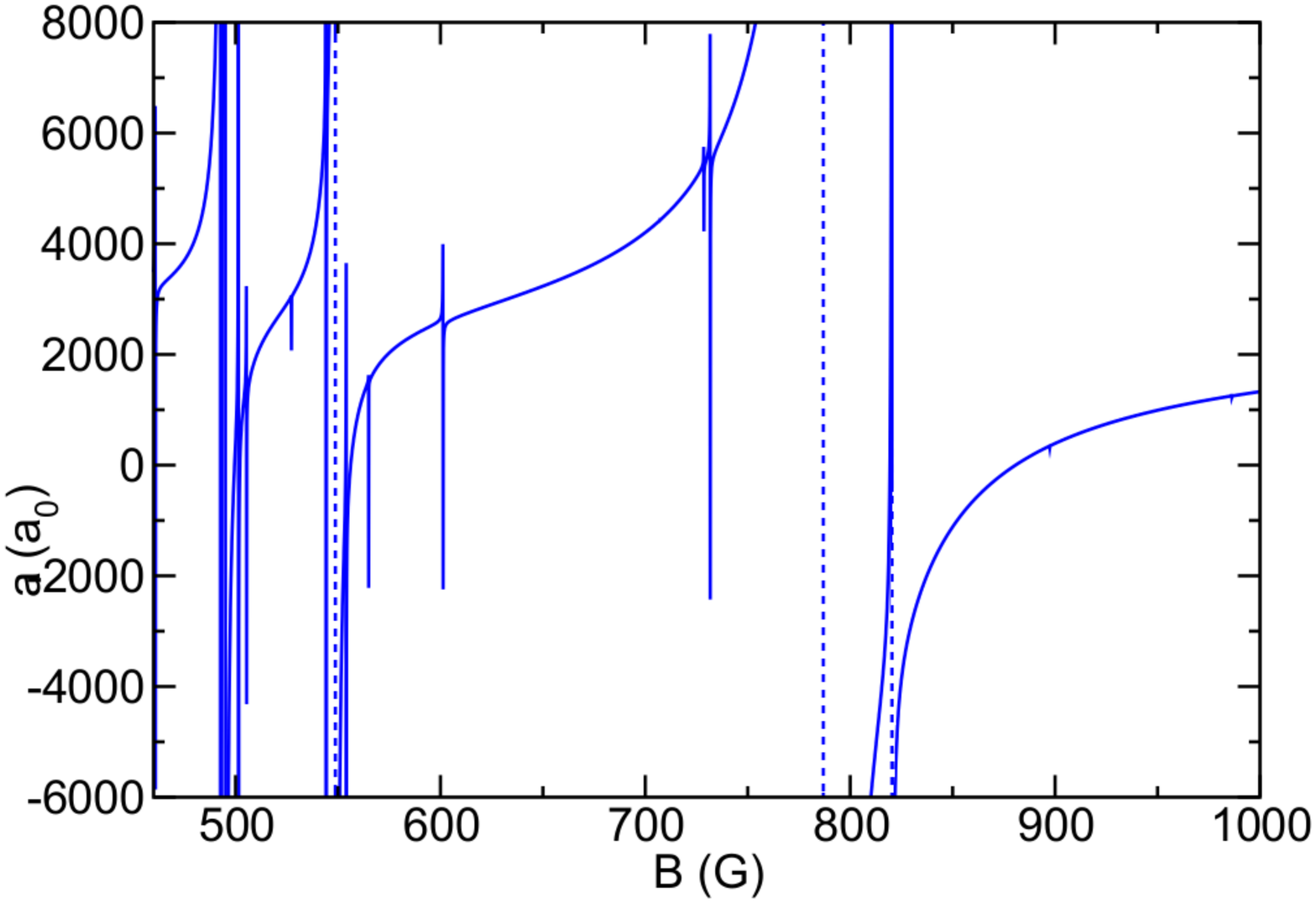}
\caption{(Color online) The scattering length between 460 G and 1000 G,
calculated for the M2012 potential with an $sdg$ basis set. The dashed
lines indicate the pole positions of the broad $s$-wave resonances near
549 G and 787 G and the $d$-wave one near 820 G.  Note that some narrow
resonances appear as spikes because of the limited grid spacing of
0.1~G, and some of the narrowest are not visible at all. }
\label{A_all}
\end{figure}

An important goal of this paper is to develop a theoretical model that is
capable of giving an accurate mapping between the scattering length $a(B)$ and
the magnetic field $B$, with particular focus on collisions between two Cs
atoms in the lowest Zeeman level of the ground-state manifold. Our new M2012
model is based on realistic potentials and includes the full spin Hamiltonian
for the Cs$_2$ molecule at long range. It has been calibrated against
experimental binding energies and a large number of resonance positions at
magnetic fields between 0 and 1000 G. We therefore expect it to provide an
accurate representation of $a(B)$ across this entire range of fields.
Figure~\ref{A_all} shows $a(B)$ on a grid with 0.1 G spacing over the new
experimental range between 460 G and 1000 G. At this resolution, some narrow
resonances are not fully resolved, and the narrowest ones are not visible at
all.

The Supplemental Material \cite{SuppMat} provides a tabulation of $a(B)$ against $B$ over the
full range of fields from 0 to 1200 G, for use in interpreting experiments.
Additional tables lists all $s$-, $d$-, $g$-, $i$- and $l$-wave resonances up
to 1000 G.

\section{Conclusion}\label{Conclusion}
We have explored the scattering properties of ground-state Cs atoms and
the binding energies of weakly bound Cs$_2$ molecules in the previously
uncharted magnetic field range up to 1000~G, using a combination of
experiment and theory. We have developed a new model of the interaction
potential that reproduces the experiments accurately over the entire
range of magnetic fields studied.

Experimentally, we have investigated Feshbach resonances and dimer
binding energies in the magnetic field range between 450~G and 1000~G,
utilizing an ultracold Cs sample in an optical dipole trap. Around
550~G and 800~G, we verified that the general scattering properties of
atomic Cs are governed by two broad $s$-wave Feshbach resonances.
Fifteen new Feshbach resonances stemming from molecular states with
$L>0$ were pinpointed by trap loss spectroscopy. We found the first
evidence for the existence of $i$-wave Feshbach resonances, resulting
from the coupling of molecular states with $L=6$ to the atomic
threshold. By performing magnetic field modulation spectroscopy, we
determined the binding energies of several dimer states, paying
particular attention to the two $s$-wave states that are responsible
for the general Cs scattering properties in the high-field region.

To calculate the scattering properties and bound-state energies, we
solve the Schr\"{o}dinger equation using coupled-channel methods. We
carried out direct least-squares fitting to the combined experimental
data of this article and Refs.\ \cite{Mark2007sou, Lange2009doa,
Zenesini2012cob, Gustavsson2008coi, Hutson:Cs2-note:2008} to obtain a new
6-parameter model of the long-range interaction potential, which we
designate M2012.

The M2012 potential reproduces the experimental results much better
than the earlier M2004 potential \cite{Chin2004pfs}, particularly at
higher fields (above 250 G). It also predicts $i$-wave and $l$-wave
states, which were not included in the least-squares fits. The
calculated positions of $l$-wave bound states agree well with the
experimental results reported in Ref.\ \cite{Knoop2008mfm}, for which
the M2004 potential failed; this demonstrates the predictive power of
the new model. The pole positions of the two broad $s$-wave Feshbach
resonances at high field are calculated to be 548.78~G and 786.80~G for
the $-6(34)7s(6)$ and the $-6(34)6s(6)$ states, respectively.

The M2012 potential does have some remaining deficiencies. In particular, it is
fitted only to results from ultracold collisions and the bound states that lies
within 10 MHz of the atomic threshold. It does not include results from
electronic spectroscopy on deeply bound levels of Cs$_2$, or the
near-dissociation levels observed in 2-color photoassociation spectroscopy
\cite{Vanhaecke:2004}, which are bound by 5 to 80 GHz. It also does not
satisfactorily reproduce the positions of $i$-wave states associated with
excited hyperfine thresholds. Resolving these remaining issues will require a
simultaneous fit to all the experiments together, and is a topic for future
work.

Our new model allows us to make an accurate connection between the
experimentally controllable magnetic field strength $B$ and the $s$-wave
scattering length $a$ over a wide range of fields. The scattering length is the
essential parameter in universal theories, and this connection was crucial in
allowing us to interpret our measurements of three-body recombination in terms
of universal Efimov physics in Ref.\,\cite{Berninger2011uot}. The present work
is important not only for experiments on Cs, but also provides important
information for ongoing and future experiments involving Cs mixtures, such as
RbCs \cite{Takekoshi2012ttp}, LiCs \cite{Repp2012ooi, Tung2012umo} and other
interesting combinations.

\begin{acknowledgments}

We acknowledge support by the Austrian Science Fund FWF within project P23106.
A. Z. was supported within the Marie Curie Intra-European Program of the
European Commission (project LatTriCs). P. S. J. and J. M. H. acknowledge support from EPSRC,
AFOSR MURI Grant FA9550-09-1-0617, and EOARD Grant FA8655-10-1-3033.

\end{acknowledgments}

\end{document}